\documentclass[12pt]{article}
\usepackage{epsfig}
\def\gsim{\;\raise0.3ex\hbox{$>$\kern-0.75em\raise-1.1ex\hbox{$\sim$}}\;}
\def\lsim{\;\raise0.3ex\hbox{$<$\kern-0.75em\raise-1.1ex\hbox{$\sim$}}\;}
\newcommand{\be}{\begin{equation}}
\newcommand{\kelvin}{^\circ \! K}
\newcommand{\ee}{\end{equation}}
\newcommand{\bea}{\begin{eqnarray}}
\newcommand{\eea}{\end{eqnarray}}
\newcommand{\bt}{\begin{tabular}}
\newcommand{\et}{\end{tabular}}
\newcommand{\ba}{\begin{array}}
\newcommand{\ea}{\end{array}}
\newcommand{\ov}{\overline}

\newcommand{\bvec}{\mathbf}
\newcommand{\smallw}{{\scriptscriptstyle W}} %
\newcommand{\smallr}{{\scriptscriptstyle R}} %
\newcommand{\smalll}{{\scriptscriptstyle L}} %
\newcommand{\gl}{g_\smalll}
\newcommand{\gr}{g_\smallr}
\newcommand{\lgr}{g_{\smalll \smallr}}
\def \dilog  {\mbox{Li}_2}
\newcommand{\slp}{/\!\!\!p}
\newcommand{\slwp}{/\!\!\!\widetilde{p}}
\def\ca{{C_{\scriptscriptstyle A}}}
\def\cv{{C_{\scriptscriptstyle V}}}
\def\tw{{\theta_{\scriptscriptstyle W}}}
\def\mw{{m_{\scriptscriptstyle W}}}
\def\mz{{m_{\scriptscriptstyle Z}}}
\begin{document}
\setlength{\unitlength}{1mm}

\setlength{\unitlength}{1mm} {\hfill
    $\ba{r}
    \mbox{DSF 21/2002} \\
    \mbox{astro-ph/0301438}
    \ea$}\vspace*{1cm}

\begin{center}
{\Large \bf Neutrino energy loss rate in a stellar plasma}
\end{center}

\bigskip\bigskip

\begin{center}
{\bf S. Esposito}, {\bf G. Mangano}, {\bf G. Miele}, {\bf I. Picardi}, and
{\bf O. Pisanti}

\vspace{.5cm}

\noindent
{\it Dipartimento di Scienze Fisiche, Universit\`{a} di Napoli ``Federico
II''\\
and \\
Istituto Nazionale di Fisica Nucleare, Sezione di Napoli \\
Complesso Universitario di Monte S. Angelo,  Via Cinthia, I-80126 Napoli,
Italy \\
E-mail: sesposito@na.infn.it, mangano@na.infn.it, miele@na.infn.it,
picardi@na.infn.it, pisanti@na.infn.it}
\end{center}

\bigskip\bigskip\bigskip

\begin{abstract}
We review the purely leptonic neutrino emission processes, contributing to
the energy loss rate of the stellar plasma. We perform a complete analysis
up to the first order in the electromagnetic coupling constant. In
particular the radiative electromagnetic corrections, at order $\alpha$,
to the process $e^+ e^- \rightarrow \nu \overline{\nu}$ at finite density
and temperature have been computed. This process gives one of the main
contributions to the cooling of stellar interior in the late stages of
star evolution. As a result of the analysis we find that the corrections
affect the energy loss rate, computed at tree level, by a factor $(-4 \div
1) \%$ in the temperature and density region where the pair annihilation
is the most efficient cooling mechanism.
\end{abstract}

\vspace*{2cm}

\begin{center}
{\it PACS number(s): 13.40.Ks, 95.30.Cq, 11.10.Wx}
\end{center}

\thispagestyle{empty}
\setcounter{page}{0}

\newpage
\baselineskip=.8cm

%---------------------------------------------------------------------------

\section{Introduction}

One of the crucial parameters which strongly affect the stellar evolution
is the cooling rate. Stars during their life can emit energy in the form
of electromagnetic or gravitational waves, and/or as a flux of neutrinos.
However, in late stages a star mainly looses energy through neutrinos, and
this is pretty independent of the mass of the star. In fact, white dwarfs
and Supernovae, which are the end points for stars with very different
masses, have both cooling rates largely dominated by neutrino production.
An accurate determination of neutrino emission rates is therefore
mandatory in order to perform a careful study of the final branches of
star evolutionary tracks. In particular, a change in the cooling rates at
the very last stages of massive star evolution could sensibly affect the
evolutionary time scale and the iron core configuration at the onset of
the Supernova explosion, whose triggering mechanism is still lacking a
full theoretical understanding \cite{janka}.

The energy loss rate due to neutrino emission (hereafter denoted by $Q$)
receives contribution from both weak nuclear reactions and purely leptonic
processes. However for the rather large values of density and temperature
which characterize the final stages of stellar evolution, the latter are
largely dominant. The leading leptonic processes are the following:
\begin{itemize}
\item[i)] pair annihilation $~~~~~~~~~~~~~~e^+ \, + \, e^- \; \rightarrow
\; \nu \, + \, \ov{\nu}$
\item[ii)] $\nu$-photoproduction $~~~~~~~~~~~~~ \gamma \, + \, e^{\pm} \;
\rightarrow \; e^{\pm} \, + \, \nu \, + \, \ov{\nu}$
\item[iii)] plasmon decay $~~~~~~~~~~~~~~~~~~~~~~~~~\gamma^\ast \;
\rightarrow \; \nu \, + \, \ov{\nu}$
\item[iv)] bremsstrahlung on nuclei $~~~~e^{\pm} \, + \, Z \; \rightarrow
\; e^{\pm} \, + \, Z \, + \, \nu \, + \, \ov{\nu}$
\end{itemize}
Each process above results to be the dominant contribution to $Q$ in
different regions of the core density--temperature plane. For very large
core temperatures, $T \gsim 10^9 \,\kelvin$, and relatively low density,
$\rho \lsim 10^5$ g cm$^{-3}$, the pair annihilation is the most efficient
cooling process. For the same values of densities but lower temperatures,
$10^8 \,\kelvin \lsim T \lsim 10^9\,\kelvin$, the $\nu$--photoproduction
gives the leading contribution. These density-temperature ranges are the
typical ones for very massive stars in their late evolution. Finally,
plasmon decay and bremsstrahlung on nuclei are mostly important for large
($\rho \gsim 10^{6}$ g cm$^{-3}$) and extremely large ($\rho \gsim 10^{9}$
g cm$^{-3}$) core densities, respectively, and temperatures of the order
of $10^8 \, \kelvin \lsim T \lsim 10^{10} \,\kelvin$. Such conditions are
typically realized in white dwarfs.

Starting from the first calculations of Ref.s
\cite{Beaudet67a,Beaudet67b}, a systematic study of the energy loss rates
for processes i)--iv) has been performed in a long series of papers
\cite{Beaudet67a}-\cite{Itoh96}. In all these analyses the pair production
rate i) has been evaluated at order $G_F^2$, i.e. at the zero--order in
the electromagnetic coupling constant $\alpha$ expansion, whereas the
remaining processes ii)--iv) are at least of order $\alpha G_F^2$. Thus to
correctly compare the energy loss rates for all processes i)--iv) it is
worth computing QED radiative corrections to pair annihilation rate i),
which may lead to a sensible change in the cooling rate $Q$. This is the
aim of our analysis, whose results have been briefly reported in Ref.
\cite{EMMPP}. In this paper we give all the details of the calculations.

The paper is organized as follows. Section 2 is devoted to a brief
summary of Born amplitude calculation for pair process, while in Section 3
we report the details of order $\alpha$ QED corrections. Neutrino
photoproduction and plasmon decay are discussed in Section 4 and 5,
respectively. Our results are summarized in Section 6.

\section{Born amplitude for pair annihilation process}

Let us consider the annihilation process $ e^-(p_1) \, + \, e^+(p_2) \;
\rightarrow\; \nu_\alpha(q_1) \, + \, \ov{\nu}_\alpha(q_2)$, where
$\alpha=e,\mu,\tau$, and the 4--momenta are defined as $p_{1,2} \equiv
(E_{1,2},{\bvec{p}}_{1,2})$ and $q_{1,2} \equiv
(\omega_{1,2},{\bvec{q}}_{1,2})$. The energy loss rate induced by this
process is obtained by integrating the squared modulus of the invariant
amplitude $M_{e^+ e^- \rightarrow \nu_\alpha \ov{\nu}_\alpha}$ over the
phase-space of the involved particles, and summing over the flavour of
final neutrinos,
\bea Q_{e^+e^-} &=& \frac{1}{(2 \pi)^6}
\int \frac{d^3 {\bvec{p}}_1}{2 E_1} \int \frac{d^3 {\bvec{p}}_2}{2 E_2}
\,(E_1 + E_2) \, F_-(E_1) \, F_+(E_2)
\nonumber \\
&\times& \left\{ \frac{1}{(2 \pi)^2}\int \frac{d^3{\bvec{q}}_1}{ 2
\omega_1} \int \frac{d^3{\bvec{q}}_2}{2 \omega_2} \,
\delta^{(4)}(p_1+p_2-q_1-q_2) \,     \sum_{{\mathrm spin,\alpha}} {|M_{e^+
e^- \rightarrow \nu_\alpha \ov{\nu}_\alpha}|}^2 \right\}\,\,\, .
\label{Qnue}
\eea
The quantities $F_{\pm}(E) = \left[ \exp \left\{ \frac{E}{T} {\pm} \xi_e
\right\} + 1 \right]^{-1}$ are the Fermi-Dirac distribution functions for
$e^{{\pm}}$ with temperature $T$ and degeneracy parameter $\xi_e$, and in
$\sum_{\mathrm spin,\alpha}$ a sum over all particle polarizations and
final flavours is performed. Notice that, as long as neutrino mean free
path is large enough that they can leave the star without any further
interaction, there is no relevant neutrino component in the stellar
plasma. Thus no neutrino distribution function is present in the
expression for the energy loss rate.

\begin{figure}
\begin{center}
\epsfysize=10cm
\epsfxsize=14cm
\epsffile{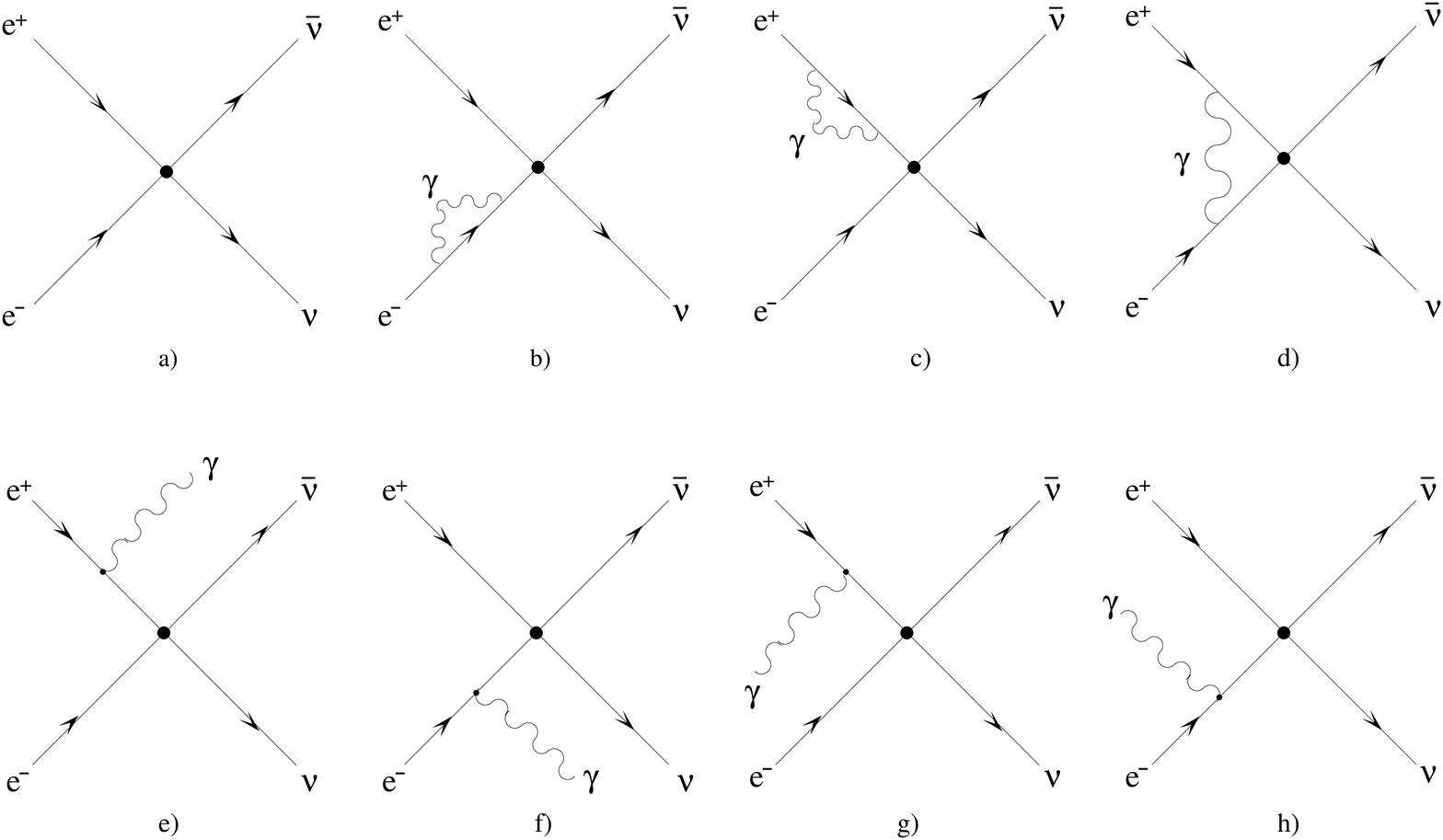}
\caption{Feynman diagrams for the pair annihilation process up to order
$\alpha G_F^2$.}
\label{pair}
\end{center}
\end{figure}

In this Section we evaluate $Q_{e^+e^-}$ in the Born approximation
(hereafter denoted with $Q_{e^+e^-}^B$), i.e. in the limit of a
four-fermion electroweak interaction and no electromagnetic radiative
correction (see Figure \ref{pair}a).
Let us consider first the pair annihilation in electron neutrinos. At
first order in perturbation theory, the diagram in Figure \ref{pair}a
contributes to the invariant amplitude $M_{e^+ e^- \rightarrow \nu_e
\ov{\nu}_e}$ with two terms coming from $W^{\pm}$ and $Z^0$ boson
exchange, respectively. By using the low energy expression for the vector
boson propagators, one has\footnote{We use natural units, $\hbar= c =
k=1$.}
\be
M^W_{e^+ e^- \rightarrow \nu_e \ov{\nu}_e}=-\frac{g^2}{8 \, \mw^2}
  [\bar{u}(q_1){\gamma}_{\mu}(1-\gamma_5)u(p_1)]~
  [\bar{v}(p_2){\gamma}^{\mu}(1-\gamma_5)v(q_2)]\,\,\, ,
\label{aw}
\ee
and
\be
M^Z_{e^+ e^- \rightarrow \nu_\alpha \ov{\nu}_\alpha}=-\frac{g^2}{8\,
 \cos^2\tw \, \mz^2}
 [\bar{v}(p_2){\gamma}_{\mu}(\cv-\ca\gamma_5)u(p_1)]~
 [\bar{u}(q_1){\gamma}^{\mu}(1-\gamma_5)v(q_2)]\,\,\, ,
\label{az}
\ee
where $\cv=2 \sin^2 \theta_\smallw - 1/2$, $\ca=-1/2$ and $\alpha = e,
\mu, \tau$. By using a Fierz transformation on Eq.(\ref{aw}) and summing
it to (\ref{az}) one gets the total amplitude in Born approximation,
\be
M^B_{e^+ e^- \rightarrow \nu_e
  \ov{\nu}_e}=-\, \frac{G_F}{\sqrt2}\, [\bar{u}(q_1){\gamma}_{\mu}(1-{\gamma}_{5})
  v(q_2)]~ [\bar{v}(p_2)\gamma^\mu (\cv'-\ca'\gamma_{5})u(p_1)]\,\,\, ,
\label{mfie}
\ee
with
\bea
\cv'&=&1+\cv=\frac{1}{2}+ 2\, {\sin}^2{\tw}\,\,\, ,\\
\ca'&=&1+\ca=\frac{1}{2}\,\,\, .
\eea
The squared modulus of the amplitude (\ref{mfie}), summed on the
polarizations of the incoming and outgoing particles, can be expressed as
the product of the two tensors, $T^{(e)}_{\mu \nu}$ and $T^{(\nu)}_{\mu
\nu}$,
\be
  \sum_{\mathrm spin}| M^B_{e^+ e^- \rightarrow \nu_e
  \ov{\nu}_e}|^2=\frac{G_F^2}{2}~ T^{(e)}_{\mu \nu}\, {T^{(\nu)}}^{\mu
  \nu}\,\,\, .
\label{sum}
\ee
The tensors $T^{(e)}_{\mu \nu}$ and $T^{(\nu)}_{\mu \nu}$ in
Eq.(\ref{sum}) can be both decomposed in a symmetric (S) and antisymmetric
(A) part in the indices $\mu\nu$, namely
\be
S^{(e,\,\nu)}_{\mu \nu} \equiv \frac{T^{(e,\,\nu)}_{\mu
\nu}+T^{(e,\,\nu)}_{\nu \mu}}{2}\,\,\, ,
\label{te}
\ee
\be
A^{(e,\,\nu)}_{\mu \nu} \equiv \frac{T^{(e,\,\nu)}_{\mu
\nu}-T^{(e,\,\nu)}_{\nu \mu}}{2}\,\,\, .
\label{tu}
\ee
In their product only the SS and AA combinations survive, but the latter
disappears after performing the integration over neutrino phase spaces.
This integration can be performed by using the Lenard formula, namely
\bea
  \int
   \frac{d^3{\bvec{q}}_1}{2\, \omega_1} \int\frac{d^3{\bvec{q}}_2}{2\, \omega_2}
   \delta^{(4)} (p-q_1-q_2){q_1}^\alpha {q_2}^\beta= \frac{\pi}{24}(2p^\alpha
   p^\beta+g^{\alpha\beta }p^2)\, \Theta(p^0)\, \Theta(p^2)\,\,\, ,
\label{lenard}
\eea
where $p^\mu$ denotes a generic 4-momentum. By means of Eq.(\ref{lenard}),
the quantity in curly brackets of Eq.(\ref{Qnue}), but only for electron
neutrinos, takes the form
\bea
&&\frac{1}{(2 \pi)^2}\int \frac{d^3{\bvec{q}}_1}{2\,\omega_1} \int
\frac{d^3{\bvec{q}}_2}{2\,\omega_2}\, \delta^{(4)}(p_1+p_2-q_1-q_2)\,
\sum_{{\mathrm spin}} {|M^B_{e^+ e^- \rightarrow \nu_e \ov{\nu}_e}|}^2
\nonumber \\
&=& \frac{8 G_F^2}{3 \pi} \,(m_e^2+p_1 \cdot p_2) \left[ \left( \cv^{\!\!\!
\prime  2} + \ca^{\!\!\! \prime  2} \right) p_1 \cdot p_2 + \left(2\,
\cv^{\!\!\! \prime  2} - \ca^{\!\!\! \prime  2} \right) m_e^2
\right]\,\,\, .
\label{Bexpr}
\eea
For $\nu_\mu$ or $\nu_\tau$ production, the $W^{\pm}$ exchange term is
instead absent and thus only the neutral current contributes. In this
case, by using the previous arguments the same expression (\ref{Bexpr}) is
obtained but with the substitution $\cv' , \ca' \rightarrow \cv , \ca$. By
virtue of the above results, the total $Q_{e^+e^-}^B$, obtained by summing
on neutrino flavour, reads
\bea
Q_{e^+e^-}^B = \frac{G_F^2\, m_e^4}{18 \pi^5} \int_0^\infty \frac{|{\bvec{p}_1}|^2
\, d |{\bvec{p}_1}|}{E_1} \int_0^\infty \frac{|{\bvec{p}_2}|^2 \, d |{\bvec{p}_2}|}
{E_2}\, \left(E_1+E_2\right) F_-(E_1) F_+(E_2) \,\,\,\,\,\,\,\,\,\,\,\,\,\,
\nonumber \\
\times \left[ \cv^{\!\!\! \prime  2} \left(\frac{4\,E_1^2
E_2^2}{m_e^4}+\frac{9\, E_1 E_2}{m_e^2} - \frac{E_1^2+E_2^2}{m_e^2} + 9
\right) + \ca^{\!\!\! \prime  2} \left( \frac{4\,E_1^2 E_2^2}{m_e^4} -
\frac{E_1^2+E_2^2}{m_e^2} \right) \right],
\nonumber \\
\label{qpair}
\eea
where we have denoted with $C_{\scriptscriptstyle V,A}^{\prime 2} \equiv
(1 + C_{\scriptscriptstyle V,A})^2 + 2 C_{\scriptscriptstyle V,A}^2$ and
we have performed the angular integrations. Note that $Q_{e^+e^-}^B$
depends on the temperature $T$ and the electron degeneracy parameter
$\xi_e$ only.

\begin{figure}
\begin{center}
\epsfysize=8cm
\epsfxsize=12cm
\epsffile{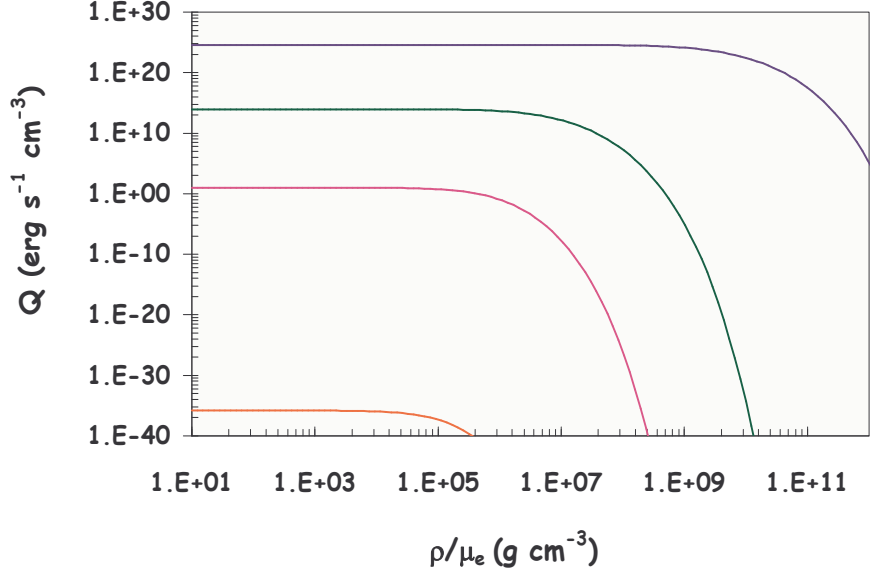}
\caption{The energy loss rate due to pair annihilation process in the Born
approximation, $Q_{e^+e^-}^B$, is here plotted for
$T=10^8,10^{8.5},10^9,10^{10} ~\kelvin$ (from bottom to top).}
\label{born}
\end{center}
\end{figure}

It is customary to recast the dependence of the energy loss rate on
$\xi_e$ (or the electron chemical potential) in terms of the matter
density, $\rho$, the temperature, $T$, and the electron molecular weight,
$\mu_e$,
\be
\frac{1}{\mu_e} \; \equiv \; \sum_i \, X_i \, \frac{Z_i}{A_i}\,\,\, ,
\ee
where in the above expression the sum is performed over all nuclides,
$Z_i$ and $A_i$ stand for the atomic and the weight number of the
$i$--nuclide, respectively, and $X_i$ is its mass fraction. To this aim,
by requiring the electrical neutrality of the plasma, we have
\be
n_{e^-} \, - \, n_{e^+} \; = \; N_A \, \frac{\rho}{\mu_e} \,\,\, ,
\label{EN}
\ee
where $n_{e^{\pm}}$ are the $e^{\pm}$ number densities and $N_A$ is the
Avogadro number. The degeneracy parameter $\xi_e$ can be then obtained by
inverting Eq.(\ref{EN}), namely
\be
\frac{\rho}{\mu_e} = \frac{1}{\pi^2 N_A} \int_{m_e}^\infty E
\sqrt{E^2-m_e^2} \, d E \, \left( \frac{1}{\exp \left\{ \frac{E}{T} -
\xi_e \right\} + 1}- \frac{1}{\exp \left\{ \frac{E}{T} + \xi_e \right\} +
1} \right)\,\,\, .
\label{rhomue}
\ee
Once performed numerically the integration over electron/positron momenta
in Eq.(\ref{qpair}), the energy loss rate due to pair annihilation in the
Born approximation can be expressed as a function of $T$ and $\rho/\mu_e$
only. The results are shown in Figure \ref{born}, where $Q_{e^+e^-}^B$ is
plotted as a function of $\rho/\mu_e$, for the following values of
temperature $T=10^8,10^{8.5},10^9,10^{10} ~\kelvin$.

\section{Radiative corrections to pair annihilation}

A consistent computation of the $\alpha G_F^2$ corrections to the energy
loss rate induced by pair annihilation is obtained by considering in
addition to the tree level graph of Figure \ref{pair}a, the {\it
radiative} diagrams of Figures \ref{pair}b-\ref{pair}h. The corrections of
order $\alpha G_F^2$ are then  obtained {\it via} their interference.

It is worth while observing that since the typical energy carried by the
outgoing neutrino pair is at most of the order of 1 MeV, one can safely
neglect the electroweak radiative corrections to the four-fermion
effective interaction (involving additional weak boson propagators), and
only consider the gauge-invariant set of purely QED contributions.

When the processes we consider take place in stellar interiors, we must
take into account the influence of the electromagnetic plasma (namely,
$e^{\pm}, \gamma$) when computing radiative corrections. To this end we
have performed calculations by using the Real Time Formalism for finite
temperature quantum field theory. In this framework the thermal
propagators for electrons and photons read as follow
\bea
i\, S(p) &=& \left( {\slp} + m \right) \left[ \frac{i}{p^2 - m_e^2 + i
\epsilon} \, - \, \Gamma_F(p)  \right]\,\,\, , \label{ProE} \\
i\, D_{\alpha \beta}(k) & = & - g_{\alpha \beta}\left[ \frac{i}{k^2 + i
\epsilon} +
\Gamma_B(k) \right] \,\,\, ,
\label{PF}
\eea
with
\bea
\Gamma_F(p) &=& 2 \pi \, \delta \left( p^2 - m_e^2 \right) \left[ F_-(|p_0|)
\Theta (p_0) + F_+(|p_0|) \Theta (-p_0) \right] \,\,\, ,\label{gf}  \\
\Gamma_B(k) &=& 2 \pi \, \delta \left( k^2  \right) B(|k_0|)\,\,\, , \label{gb}
\eea
where $\Theta (x)$ is the step function and $B(x)$ is the Bose-Einstein
distribution function. The first terms in Eqs.(\ref{ProE}) and (\ref{PF})
are the usual $T=0$ Feynman propagators, while those depending on the
temperature (and density) through the distribution functions describe the
interactions with real particles of the thermal bath.

According to the different diagrams contributing to radiative corrections,
it is customary to classify these corrections as follows:
\begin{itemize}
\item[-] electron mass and wavefunction renormalization (Figures
\ref{pair}b--\ref{pair}c);
\item[-] electromagnetic vertex correction (Figure \ref{pair}d);
\item[-] $\gamma$ emission/absorption (Figures
\ref{pair}e--\ref{pair}h).
\end{itemize}

\subsection{Zero--temperature radiative correction}
\label{vacuum}

By using in the evaluation of radiative corrections of Figures
\ref{pair}b-\ref{pair}d the first term only in the propagators of
Eqs.(\ref{ProE}),(\ref{PF}), one obtains the off-shell contribution to the
radiative amplitude, and the corresponding result is known as
zero--temperature or vacuum correction.

This issue has been addressed in Ref. \cite{Passera} for the process $e^-
\, + \, \nu \rightarrow e^- \, + \, \nu$, but we can easily obtain the
desired corrections for pair annihilation by using crossing symmetry. The
vacuum correction to be inserted into Eq.(\ref{Qnue}) is then
\be
\sum_{{\mathrm spin},\alpha} {|\Delta M^{T=0}_{e^+ e^- \rightarrow
\nu_\alpha \ov{\nu}_\alpha}|}^2 \; = \; \frac{128}{\pi} \, \alpha\, G_F^2
\, m_e^2 \, \omega_2^2 \, \delta(E_2,\omega_2)\,\,\, ,
\label{pas1}
\ee
where
\bea
\delta(E_2,\omega_2) &=& \gl^2\; \left\{V_1(E_2) +V_2(E_2)\left[
z-1+\frac{m_e\, z}{2\, \omega_2} \right] \right\} \nonumber  \\
&+& \gr^2\; \left\{V_1(E_2)\left(1-z\right)^2 + V_2(E_2)\left[ z - 1 +
\frac{m_e\, z}{2\, \omega_2} \right] \right\} \nonumber \\
&+& \lgr \,\left\{ \left[V_1(E_2)-V_2(E_2)\right] \left(\frac{m_e\,
z}{\omega_2} \right) -2\, V_2(E_2) \left[z-1-z^2 \right] \right\}\,\,\, ,
\eea
\bea V_1(E_2) = {\mathrm Re} \left\{ \left( 2\, \log \frac{m_e}{\lambda}
\right) \left[1-\frac{E_2}{2\, |{\bvec{p}_2}|} \log \left( \frac{E_2 +
|{\bvec{p}_2}|} {E_2 - |{\bvec{p}_2}|} \right) \right] -2 -
\frac{E_2}{|{\bvec{p}_2}|} \left[ \dilog \left( \frac{|{\bvec{p}_2}| - E_2
- m_e}{2\, |{\bvec{p}_2}|}
\right) \right. \right. \nonumber \\
- \left. \left. \dilog \left( \frac{|{\bvec{p}_2}| + E_2 + m_e} {2\,
|{\bvec{p}_2}|} \right) \right] - \frac{1}{4 |{\bvec{p}_2}|} \left[ -3 E_2
+ m_e + E_2 \log \left( \frac{2\, m_e - 2\, E_2}{m_e} \right) \right] \log
\left( \frac{E_2 + |{\bvec{p}_2}|}
{E_2-|{\bvec{p}_2}|} \right) \right\}, \nonumber \\
\eea
\bea
V_2(E_2) =- \frac{m_e}{4\, |{\bvec{p}_2}|} \log \left( \frac{E_2 +
|{\bvec{p}_2}|}{E_2 - |{\bvec{p}_2}|}\right)\,\,\,,
\eea
with
\bea
\gl^2 &=& 3 \sin^4 \theta_\smallw - \sin^2 \theta_\smallw + \frac{3}{4}
\,\,\, , \\
\gr^2 &=& 3 \sin^4 \theta_\smallw \,\,\,,\\
\lgr  &=& \sin^2 \theta_\smallw \left( 3 \sin^2 \theta_\smallw -
\frac{1}{2} \right)\,\,\, ,
\eea
$z=(E_2+m_e)/\omega_2$ and $\lambda$ is a small photon mass introduced to
regularize the infrared divergences. Remind that the dilogarithm function
$\dilog(x)$ is defined by
\be
\dilog(x) \equiv -\int_0^x \!dt \,\frac{\log(1-t)}{t}\,\,\,.
\ee
The expression (\ref{pas1}) depends on the infrared regulator $\lambda$
and diverges for $\lambda \rightarrow 0$. This is quite obvious since as
it is well known, when computing QED radiative corrections, in addition to
the pair annihilation process one must consider also the bremsstrahlung
radiation accompanying the process (Figures \ref{pair}e-\ref{pair}h). In
fact, only the combination of virtual and real photon corrections is free
from infrared divergencies.

The bremsstrahlung process will be considered in detail in Section
\ref{bremcor} but here we anticipate the strategy for a careful and
reliable numerical computation of the integrals in Eq.(\ref{dqt0}). In
fact, numerical cancellation of divergencies is very hard to handle and we
prefer to divide the bremsstrahlung contribution into a ``soft'' part
(according to the photon energy being lower than some fixed threshold
$\epsilon$) which will cancel the infrared divergencies of the pair
process, and a ``hard'' one which, in our case, will depend on the thermal
photon distribution function. The last one will be considered in Section
\ref{bremcor} while we now report the Soft-Bremsstrahlung (S.B.) squared
modulus. Again we use the result of Ref. \cite{Passera} with a suitable
crossing transformation and thus we get
\be
\sum_{{\mathrm spin},\alpha} {|\Delta M^{S.B.}_{e^+ e^- \rightarrow
\nu_\alpha \ov{\nu}_\alpha}|}^2 \; = \; \frac{128}{\pi}\, \alpha\, G_F^2
\, m_e^2 \, \omega_2^2 \, I_\gamma(E_2,\epsilon) \,  \left[\gl^2 +\gr^2
\left(1- z\right)^2 +\lgr  \left(\frac{m_e
z}{\omega_2}\right)\right]\,\,\, ,
\label{pas2}
\ee
where
\bea
&& I_\gamma(E_2,\epsilon) = {\mathrm Re} \left\{ \left( 2\, \log
\frac{\lambda}{\epsilon} \right) \left[ 1 - \frac{E_2}{2\, |{\bvec{p}_2}|}
\log \left( \frac{E_2 + |{\bvec{p}_2}|}{E_2 - |{\bvec{p}_2}|} \right)
\right] + \frac{E_2}{2\, |{\bvec{p}_2}|} \left[\, L \left( \frac{E_2 +
|{\bvec{p}_2}|}{E_2 - |{\bvec{p}_2}|} \right) \right. \right. \nonumber \\
&&- \left. \left. L \left( \frac{E_2 - |{\bvec{p}_2}|}{E_2 + |{\bvec{p}_2}|}
\right) + \log \left( \frac{E_2 + |{\bvec{p}_2}|}{E_2 - |{\bvec{p}_2}|}
\right) \left( 1 - 2\, \log \left( \frac{|{\bvec{p}_2}|}{m_e} \right)
\right) \right] +1-2\, \log 2 \right\}\,\,\, ,
\eea
and
\be
L(x) \equiv  \int_0^x \!dt \,\frac{\log|1-t|}{t}\,\,\, .
\ee
Remarkably, by summing the expressions (\ref{pas1}) and (\ref{pas2}),
since the term $[V_1(E_2)+I_\gamma(E_2,\epsilon)]$ does not depend on the
infrared regulator $\lambda$, the total squared amplitude is now infrared
divergence free. Then, the term in curly brackets in Eq.(\ref{Qnue}), once
integrating over all variables but one by using the $\delta$--function,
becomes:
\bea
\Phi & \equiv & \frac{1}{(2 \pi)^2}\int \frac{d^3{\bvec{q}}_1}{ 2
\omega_1} \int \frac{d^3{\bvec{q}}_2}{2 \omega_2} \,
\delta^{(4)}(p_1+p_2-q_1-q_2) \nonumber \\
&\times& \sum_{{\mathrm spin},\alpha} \left( {|\Delta M^{T=0}_{e^+ e^-
\rightarrow \nu_\alpha \ov{\nu}_\alpha}|}^2 + {|\Delta M^{S.B.}_{e^+ e^-
\rightarrow \nu_\alpha \ov{\nu}_\alpha}|}^2 \right) \nonumber \\
& = &\frac{16\, \alpha\, G_F^2\, m_e^2}{\pi^2} \int_{\omega_m}^{\omega_M}
d \omega_2^\prime \, \frac{\omega_2^{\prime 2}}{|{\bvec{p}_2^\prime}|}
\nonumber \\
&\times& \left\{ \delta(E_2^\prime,\omega_2^\prime) + I_\gamma(E_2^\prime,
\epsilon)\, \left[\gl^2 +\gr^2 \left(1- z^\prime\right)^2 +\lgr
\left(\frac{m_e z^\prime}{\omega_2}\right)\right]\right\} \,\,\, ,
\eea
where we denote with a prime the quantities in the electron rest frame,
and the integration limits are
\be
\omega_m \, = \, m_e \, \frac{E_2^\prime + m_e}{E_2^\prime +
|{\bvec{p}_2^\prime}| + m_e}~~~, ~~~ \omega_M \, = \, m_e \,
\frac{E_2^\prime + m_e}{E_2^\prime - |{\bvec{p}_2^\prime}| +
m_e}\,\,\, .
\label{inteom}
\ee
Using this result, the correction to the energy loss rate at
zero--temperature is
\be
\Delta Q^{T=0}_{e^+e^-} \; = \; \frac{1}{32\,\pi^4}\,
\int_0^\infty d |{\bvec{p}_1}| \, d |{\bvec{p}_2}|\, \int_{-1}^{1}
d (\cos \theta_{12})\, |{\bvec{p}_1}|^2 \, |{\bvec{p}_2}|^2 \,
\frac{E_1 + E_2}{E_1\,E_2}\, \hat{\Phi}\, F_-(E_1)\, F_+
(E_2)\,\,\, ,
\label{dqt0}
\ee
where $\theta_{12}$ is the angle between ${\bvec p}_1$ and ${\bvec p}_2$,
$\hat{\Phi}$ stands for $\Phi$ boosted to the comoving frame and the
integrals must be numerically evaluated.

\subsection{Thermal radiative corrections}

The {\it true} thermal radiative corrections to the Born estimate of the
neutrino pair production process come from considering in the evaluation
of radiative diagrams of Figures \ref{pair}b-\ref{pair}d the terms
involving at least one thermal part of the propagators (\ref{ProE}) and
(\ref{PF}). These contributions do not involve ultraviolet divergencies
due to the presence of the Fermi and Bose function in the thermal
propagators. Nevertheless, the results corresponding to each of these
corrections are not free from infrared divergencies and, in general, are
gauge-dependent too. The inclusion in our calculations of the photon
emission/absorption diagrams in Figures \ref{pair}e-\ref{pair}h is then
required to overcome these difficulties, as we have directly and carefully
checked. In the intermediate steps, however, we have to regularize the
actual divergences and, to this end, we have explicitly subtracted all
divergent terms expanding the squared amplitudes in a Laurent series
around the pole singularities. This method follows quite closely what has
been already used in Refs. \cite{EMMP98}, \cite{EMMP99}.

\subsubsection{Mass and wavefunction renormalization at finite temperature}

As already stated, at order $\alpha$, the $e^{\pm}$ thermal mass shift and
thermal wavefunction renormalization corrections come from the
interference of the diagrams in Figures \ref{pair}b-\ref{pair}c with the
tree-level one of Figure \ref{pair}a, having subtracted the
zero--temperature contribution.

The thermal mass correction to neutrino energy loss rate may be obtained
by replacing $e^{\pm}$ mass $m_e$ with the renormalized value
$m_{e^{\pm}}^R = m_e + \delta m_{{\pm}}$ in the Born expression for the
rate and subtracting the zero--temperature limit (\ref{qpair}). The
self-energy for an electron of energy $E$ and momentum ${\bf p}$ has been
calculated in Ref. \cite{EMMP99} and we refer the reader to this paper for
details. At order $\alpha$ the thermal mass shift for $e^{\pm}$ is given
by
\bea
\delta m_{\pm} &=& \frac{\alpha \pi}{3 m_e} \, T^2 + \frac{\alpha}{\pi
m_e} \int_0^\infty \, d|{\bvec{k}}| \, \frac{|{\bvec{k}}|^2}{E_k} \,
\left( F_{\pm}(E_k) + F_\mp(E_k) \right) \nonumber \\
& + & \, \frac{\alpha m_e}{2 \pi |{\bvec{p}}|} \, \int_0^\infty \,
d|{\bvec{k}}| \frac{|{\bvec{k}}|}{E_k} \, \left( F_{\pm}(E_k) \, \log C_-
+ F_\mp(E_k) \, \log C_+ \right)\,\,\, ,
\label{dme}
\eea
with
\be
C_{\pm} \; = \; \frac{{{m_e}}^2  +  |{\bvec{p}}|\, |{\bvec{k}}|\, {\pm} \ E
\,{E_k}} {{{m_e}}^2 - |{\bvec{p}}|\, |{\bvec{k}}|{\pm} \ E\,{E_k}} \,\,\, ,
\label{cpm}
\ee
where $E_k=\sqrt{|{\bvec{k}}|^2 + m_e^2}$ and $E=\sqrt{|{\bvec{p}}|^2 +
m_e^2}$. By using the thermal renormalized masses in the expressions for
energies appearing in the Born rate in Eq.(\ref{qpair}), after some
algebra we obtain the following thermal mass correction to the neutrino
energy loss rate
\be
\Delta Q^M_{e^+e^-} \; = \; \Delta Q^{M,B}_{e^+e^-} + \Delta Q^{M,
{F_+}}_{e^+e^-} + \Delta Q^{M,{F_-}}_{e^+e^-}\,\,\, ,
\label{dqmass}
\ee
with
\bea
\Delta Q^{M,B}_{e^+e^-}  &=& \frac{\alpha G_F^2}{9 \pi^6} \,
\int_0^\infty\, d |{\bvec{p}_1}| d |{\bvec{p}_2}| d |{\bvec{k}}|\,
|{\bvec{p}_1}|^2 |{\bvec{p}_2}|^2 |{\bvec{k}}| \left( \frac{1}{E_1} +
\frac{1}{E_2} \right) \nonumber \\
&\times& F_-(E_1) F_+(E_2) B(|{\bvec{k}}|) \left( \frac{f_1}{E_1} +
\frac{f_2}{E_2} \right)\,\,\, ,
\eea
\bea
\Delta Q^{M,{F_{\pm}}}_{e^+e^-} &=& \frac{\alpha G_F^2}{18 \pi^6} \,
\int_0^\infty \, d |{\bvec{p}_1}| d |{\bvec{p}_2}| d |{\bvec{k}}|\,
\frac{|{\bvec{p}_1}|^2 |{\bvec{p}_2}|^2 |{\bvec{k}}|}{E_k} \left(
\frac{1}{E_1} +  \frac{1}{E_2} \right) F_-(E_1) F_+(E_2) F_{\pm}(E_k)
\nonumber \\
&\times& \left[ |{\bvec{k}}| \left( \frac{f_1}{E_1} +  \frac{f_2}{E_2}
\right) + \frac{m_e^2}{2} \left( \frac{f_1}{E_1 |{\bvec{p}_1}|} \log
C_{{\pm} 1} + \frac{f_2}{E_2 |{\bvec{p}_2}|} \log C_{\mp 2} \right)
\right]\,\,\, ,
\eea
\bea
f_1 &=& \frac{E_2}{m_e} \left[ (\ca^{\!\!\! \prime  2} + \cv^{\!\!\!
\prime 2}) \left( 3 + 6 \frac{E_1 E_2}{m_e^2} \right) - 3 (\ca^{\!\!\!
\prime  2} - 2 \cv^{\!\!\! \prime  2}) \right] - m_e \left( \frac{E_2}
{E_1(E_1+E_2)}\nonumber \right. \\
&+& \left. \frac{1 - F_-(E_1)}{T} \right) \left\{ (\ca^{\!\!\! \prime  2} +
\cv^{\!\!\! \prime  2}) \left[ 3 \frac{E_1 E_2}{m_e^2} \left( 1 +
\frac{E_1 E_2}{m_e^2} \right) + \frac{|{\bvec{p}_1}|^2 |{\bvec{p}_2}|^2}{m_e^4}
\right] \right. \nonumber \\
&-& \left.  3 (\ca^{\!\!\! \prime  2}- 2 \cv^{\!\!\! \prime  2}) \left( 1 +
\frac{E_1 E_2}{m_e^2} \right) \right\}\,\,\, , \\
f_2 &=& \frac{E_1}{m_e} \left[ (\ca^{\!\!\! \prime  2} + \cv^{\!\!\!
\prime 2}) \left( 3 + 6 \frac{E_1 E_2}{m_e^2} \right) - 3 (\ca^{\!\!\!
\prime  2} - 2 \cv^{\!\!\! \prime  2}) \right] - m_e \left(
\frac{E_1}{E_2(E_1 + E_2)} \nonumber \right. \\
&+&\left. \frac{1 - F_+(E_2)}{T} \right) \left\{ (\ca^{\!\!\! \prime  2}
+ \cv^{\!\!\! \prime  2}) \left[ 3 \frac{E_1 E_2}{m_e^2} \left( 1 +
\frac{E_1 E_2}{m_e^2} \right) + \frac{|{\bvec{p}_1}|^2 |{\bvec{p}_2}|^2}
{m_e^4} \right] \right. \nonumber \\
&-& \left.  3 (\ca^{\!\!\! \prime  2}- 2 \cv^{\!\!\! \prime  2}) \left( 1 +
\frac{E_1 E_2}{m_e^2} \right) \right\}\,\,\, ,
\eea
where $C_{{\pm}1,2}$ are obtained from $C_{\pm}$ in Eq.(\ref{cpm}) with
the substitution $(E,{\bvec{p}}) \rightarrow (E_{1,2}, {\bvec{p}_{1,2}})$.
Note that in Eq.(\ref{dqmass}) we have subtracted all divergent terms as
described above.

Diagrams in Figures \ref{pair}b--\ref{pair}c are also responsible for the
thermal wavefunction renormalization correction which, in the calculation
for the neutrino energy loss rate, can be obtained by using a thermal
renormalized projector on positive/negative energy states as described in
Refs. \cite{EMMP98, EMMP99}. The projector on positive energy states for
an electron of 4-momentum $p\equiv(E,{\bvec{p}})$ can be cast in the
following form
\bea
\Lambda_R^+ &=& \frac{\slp + m_e^R + \delta_+}{2 E}\,\,\, ,  \label{prop}
\\
\delta_+ &=& (\slp+m_e ) {\cal A} + \left(
\slwp-\frac{|{\bvec{p}}|^2}{m_e} \right) {\cal B}\,\,\, ,
\eea
with $\widetilde{p}=(0,{\bvec{p}})$ and where the momentum-dependent
functions ${\cal A}$ and ${\cal B}$ are given by
\bea
{\cal A} &=& \frac{\alpha}{2 \pi E} \, \int_0^\infty d|{\bvec{k}}| \,
|{\bvec{k}}|\, \left[ -\frac{2 B(|{\bvec{k}}|)}{|{\bvec{p}}|} \log \left(
\frac{E + |{\bvec{p}}|}{E - |{\bvec{p}}|}\right) + F_-(E_k)\, \left(
\frac{2\, |{\bvec{k}}|} {{\left( E - E_k \right) }^2} + \frac{\log C_-}
{|{\bvec{p}}|} \right) \right. \nonumber \\
&-& \left. F_+(E_k)\,  \left( \frac{2\,|{\bvec{k}}|}{{\left( E + E_k
\right) }^2} + \frac{\log C_+}{|{\bvec{p}}|} \right) \right]\,\,\, ,
\label{defa} \\
{\cal B} &=& \frac{\alpha}{2 \pi |{\bvec{p}}|^3 E} \, \int_0^\infty
d|{\bvec{k}}|\, \frac{|{\bvec{k}}|}{E_k} \left[ E_k \left( 4 |{\bvec{p}}|
E - 2\, m_e^2 \log \left(\frac{E + |{\bvec{p}}|}{E - |{\bvec{p}}|}\right)
\right) B(|{\bvec{k}}|) + \left( 2 |{\bvec{k}}| |{\bvec{p}}| E \right.
\right. \nonumber \\
&-& \left. \left.  m_e^2 (E-E_k) \log C_- \right) F_-(E_k) + \left( 2
|{\bvec{k}}| |{\bvec{p}}| E - m_e^2 (E+E_k) \log C_+ \right) F_+(E_k)
\right],
\label{defb}
\eea
The projector on negative energy states reads instead
\bea
\Lambda_R^- & = & \frac{\slp - m_e^R + \delta_-}{2 E}\,\,\, ,  \label{pron} \\
\delta_- &=& (\slp-m_e ) {\cal A} + \left(
\slwp+\frac{|{\bvec{p}}|^2}{m_e} \right) {\cal B}\,\,\, ,
\eea
and $\hat{{\cal A}}$ and $\hat{{\cal B}}$ can be obtained from ${\cal A}$
and ${\cal B}$ by replacing $\xi_e \rightarrow -\xi_e$. The contribution
due to $e^{\pm}$ thermal wavefunction renormalization is then obtained by
using the thermal renormalized projectors $\Lambda^{\pm}_R$ in the
evaluation of the Born rate. This procedure, having subtracted the
contribution due to the mass renormalization, gives at order $\alpha$
\bea
\Delta Q^W_{e^+e^-} &=& \frac{G_F^2}{12 \pi^5}\, \int_0^\infty d
|{\bvec{p}_1}| d |{\bvec{p}_2}| \, |{\bvec{p}_1}|^2 |{\bvec{p}_2}|^2
\left( \frac{1}{E_1} +
\frac{1}{E_2} \right) \, F_-(E_1) F_+(E_2) \nonumber \\
&& \!\!\!\!\!\!\!\!\!\!\!\!\!\!\!\!\!\!\!\!\!\!\!\!\!\!\!\!\!\!\!\!\! \times
\left( \left\{ \left( \ca^{\!\!\! \prime  2} + \cv^{\!\!\! \prime 2}
\right) \left[ \left( E_1 E_2 + m_e^2 \right) \left( 2 E_1 E_2 +m_e^2
\right) + \frac{2}{3}\, |{\bvec{p}_1}|^2 |{\bvec{p}_2}|^2 \right]
\right\} ({\cal A}_1+\hat{{\cal A}}_2) \right. \\
&& \!\!\!\!\!\!\!\!\!\!\!\!\!\!\!\!\!\!\!\!\!\!\!\!\!\!\!\!\!\!\!\!\! +
\left. \frac{2}{3} \left( \ca^{\!\!\! \prime  2} + \cv^{\!\!\! \prime 2}
\right) |{\bvec{p}_1}|^2 |{\bvec{p}_2}|^2 ({\cal B}_1+\hat{{\cal B}}_2) -2
\left( 2\, \cv^{\!\!\! \prime  2} - \ca^{\!\!\! \prime 2} \right) \left(
E_1 E_2 + m_e^2 \right) (|{\bvec{p}_1}|^2 {\cal B}_1+|{\bvec{p}_2}|^2
\hat{{\cal B}}_2) \right)\,\,\, . \nonumber
\label{dqwave}
\eea
The functions ${\cal A}_{1,2}$, ${\cal B}_{1,2}$, $\hat{{\cal A}}_{1,2}$,
$\hat{{\cal B}}_{1,2}$ are obtained from ${\cal A}$, ${\cal B}$,
$\hat{{\cal A}}$, $\hat{{\cal B}}$ with the substitution $(E,{\bvec{p}})
\rightarrow (E_{1,2}, {\bvec{p}_{1,2}})$. Again, in the numerical
computation of the integrals appearing in the expression for $\Delta
Q^W_{e^+e^-}$ we have subtracted all divergent terms.

\subsubsection{Vertex renormalization at finite temperature}
\label{vertcor}

The lowest order electromagnetic vertex correction to the neutrino energy
loss is provided by the interference term between the diagram in Figure
\ref{pair}d and the tree amplitude of Figure \ref{pair}a. In the amplitude
of the vertex diagram, three particle propagators appear, each of them
consisting of two terms according to Eqs.(\ref{ProE}) and (\ref{PF}).
However, the computation of the thermal vertex renormalization correction
to the rate is simplified by 4-momentum conservation arguments. In fact,
the mentioned three propagators would produce eight terms in the
amplitude, one of which is a pure vacuum term already considered in Sect.
\ref{vacuum}. Three of the remaining terms, proportional to a Fermi-Dirac
distribution function times a Bose-Einstein distribution function, give no
contribution to the rate since they correspond to an electromagnetic
vertex with a real photon and two real massive particles, which is not
allowed by 4-momentum conservation. Moreover, the term proportional to two
Fermi functions does not contribute to the interference term with the tree
level amplitude since it is purely imaginary. Therefore, we are left with
only three terms, one proportional to a Bose-Einstein function and the
others proportional to a Fermi-Dirac distribution
\bea
\Delta Q^V_{e^+e^-} = -\frac{\alpha G_F^2}{48 (2 \pi)^{10}} \, \int d^3
{\bvec{p}}_1 d^3 {\bvec{p}}_2 d^4 k \, \left( \frac{1}{E_1} +
\frac{1}{E_2} \right) F_-(E_1) F_+(E_2) \, \Phi_V \, \nonumber \\
\times \left[ \frac{\Gamma_B(k)}{[(k+p_1)^2 - m_e^2] [(k-p_2)^2 - m_e^2]}
- \frac{\Gamma_F(k+p_1)}{[(k-p_2)^2 - m_e^2] k^2} -
\frac{\Gamma_F(k-p_2)}{[(k+p_1)^2 - m_e^2] k^2} \right]
\label{dqvtm}
\eea
where
\bea
\Phi_V &=& 512 \, \cv^{\!\!\! \prime  2} \left[ \left( m_e^2 + p_1 \cdot
p_2 \right)\, \left( -2\,p_1 \cdot p_2\, \left( 2 m_e^2 + p_1 \cdot p_2
\right) - k \cdot p_2\,\left( 3 m_e^2 + 2\,
p_1 \cdot p_2 \right) \right. \right. \nonumber \\
&+& \left. \left. k \cdot p_1\,\left( 3 m_e^2 + 2\, k \cdot p_2 +
2\, p_1 \cdot p_2 \right) \right) \right] \nonumber \\
&-&  512 \, \ca^{\!\!\! \prime  2} \left[ m_e^2 {\left( k \cdot p_1
\right)}^2 + m_e^2 {\left( k \cdot p_2 \right) }^2 + 2\, k \cdot p_2\, p_1
\cdot p_2\, \left( m_e^2 + p_1 \cdot p_2 \right)
\right. \nonumber \\
&-&
\left. 2\,k \cdot p_1\, p_1 \cdot p_2\, \left( m_e^2 + k \cdot p_2 + p_1
\cdot p_2 \right) + 2\, p_1 \cdot p_2\, \left( -m_e^4 + {\left( p_1 \cdot
p_2 \right) }^2 \right) \right]\,\,\, .
\eea
The infrared-safe part of these terms (in the limit $k \rightarrow 0$) is
obtained by expanding the matrix element function $\Phi_V$ in powers of
$k$.

For the Bose-Einstein term (the first one in square brackets in
Eq.(\ref{dqvtm})), due to the presence of the function $B(|k_0|)$, one has
to subtract from $\Phi_V$ the terms in powers of $k$ of order 0 and 1. By
choosing the frame where the momentum $\bvec{k}$ lies along the $z$-axis
and ${\bvec{p}}_1$ is in the $x$-$z$ plane, denoting with $\theta_{1,2}$
the angle between $\bvec{k}$ and ${\bvec{p}}_{1,2}$ and with $\phi$ the
azimuthal angle of the vector ${\bvec{p}}_2$, after some algebra one gets
\bea
\Delta Q^{V,B}_{e^+e^-} &= & \frac{\alpha G_F^2}{6144 \, \pi^7} \,
\int_0^\infty d |{\bvec{p}_1}| d |{\bvec{p}_2}| d |{\bvec{k}}|
\int_{-1}^{1} d x_1 d x_2\, |{\bvec{p}_1}|^2 |{\bvec{p}_1}|^2 |{\bvec{k}}|
\left( \frac{1}{E_1} + \frac{1}{E_2} \right) \nonumber \\
&\times& F_-(E_1) F_+(E_2) B(|{\bvec{k}}|)\, I_B\,\,\, ,
\label{dqbv}
\eea
where $x_{1,2}=\cos \theta_{1,2}$ and $I_B$ is reported in Appendix
\ref{vertapp}.

The fermionic part (the second and third terms in square brackets in
Eq.(\ref{dqvtm})) contains, instead, the expression
\be
d^4 k \, \frac{\delta (k^2 + 2 k \cdot p_1) }{(k^2 - 2 k \cdot p_2) k^2}\,
\Phi_V \,\,\, ,
\label{tfdiv1}
\ee
which, assuming that the integration in $d k_0$ does not introduce
divergent terms, is proportional to
\be
d^3 {\bvec{k}} \, \frac{\Phi_V}{k \cdot (p_1+p_2)~ k \cdot p_1}\,\,\, .
\label{tfdiv2}
\ee
By using the properties of the $\delta$-function, for $k \rightarrow 0$ we
can write
\bea
k_0 &=& - E_1 + \sqrt{E_1^2 + |{\bvec{k}}|^2 + 2 |{\bvec{k}}|
|{\bvec{p}_1}| x_1} \nonumber \\
& \simeq & \frac{|{\bvec{p}_1}| x_1}{E_1} \, |{\bvec{k}}| + \frac{E_1^2 -
|{\bvec{p}_1}|^2 x_1}{2 E_1^3}\, |{\bvec{k}}|^2 + O \left( |{\bvec{k}}|^3
\right)\,\, ,
\label{tfdiv3}
\eea
and substituting in the expression in (\ref{tfdiv2}) we find that
\be
d^4 k \, \frac{\delta (k^2 + 2 k \cdot p_1) }{(k^2 - 2 k \cdot p_2) k^2}\,
\Phi_V \sim d |{\bvec{k}}| \, \frac{\Phi_V}{|{\bvec{k}}|}\,\,\, .
\label{tfdiv4}
\ee
Thus, the infrared-safe part of the fermionic term is obtained by
subtracting from $\Phi_V$ the terms in powers of $k$ of order 0 only. With
the same notations as above, after some algebra we get
\bea
\Delta Q^{V,F}_{e^+e^-} &=& \frac{\alpha G_F^2}{6144 \pi^7} \,
\int_0^\infty d |{\bvec{p}_1}| d |{\bvec{p}_2}| d |{\bvec{k}}|
\int_{-1}^{1} d x_1 d x_2\, \frac{|{\bvec{p}_1}|^2 |{\bvec{p}_2}|^2
|{\bvec{k}}|^2}{E_k} \left( \frac{1}{E_1} + \frac{1}{E_2} \right)\, \nonumber \\
&\times& F_-(E_1) F_+(E_2) \, \left[ I_{F_1}\, F_-(E_k) + I_{F_2}\, F_+(E_k)
\right]\,\,\, ,
\label{dqfv}
\eea
where $E_k=\sqrt{|{\bvec{k}}|^2+m_e^2}$. The functions $I_{F_{1,2}}$ are
defined as follows
\bea
I_{F_1} = \int_0^{2 \pi} \!\!\!\! d \phi \, \left[ \left.
\frac{\tilde{\Phi}_V^{IS}} {[(k-p_1-p_2)^2 - m_e^2] (k-p_1)^2}
\right|_{k_0=E_k} \!\!\!\!\!\!\!\!\! + \left. \frac{\tilde{\Phi}_V^{IS}}
{[(k-p_1-p_2)^2 - m_e^2] (k-p_1)^2} \right|_{k_0=-E_k} \right], \nonumber \\
\label{if1} \\
I_{F_2} = \int_0^{2 \pi} \!\!\!\! d \phi \, \left[ \left.
\frac{\hat{\Phi}_V^{IS}} {[(k+p_1+p_2)^2 - m_e^2] (k+p_2)^2}
\right|_{k_0=E_k} \!\!\!\!\!\!\!\!\! + \left. \frac{\hat{\Phi}_V^{IS}}
{[(k+p_1+p_2)^2 - m_e^2] (k+p_2)^2} \right|_{k_0=-E_k} \right], \nonumber \\
\label{if2}
\eea
where $\tilde{\Phi}_V^{IS}$ ($\hat{\Phi}_V^{IS}$) is obtained from
$\Phi_V$ subtracting the terms of order 0 in $k$ and making the
substitution $k\rightarrow k-p_1$ ($k\rightarrow k+p_2$). The integrations
in Eqs.(\ref{if1}), (\ref{if2}) may be performed analytically, since they
involve integrals of the form
\be
I \; = \; \int_0^{2 \pi} d \phi\, \frac{{n_1} + n_2\, \cos\phi + n_3\,
\cos^2\phi + n_4\, \cos^3\phi} {d_1 - d_2\, \cos\phi}\,\,\, ,
\label{ii}
\ee
where $n_i$, $d_i$ do not depend on $\phi$. However, since the denominator
above may vanish for given values of the momenta involved, the result of
the integrations depends on the kinematical region one considers. For each
function $I_{F_1}$ and $I_{F_2}$ we distinguish three different cases:
\begin{enumerate}
\item $d_2=0$: \\
the integrand functions are just polynomials in $\cos(\phi)$ and no pole
is involved:
\be
I \; = \; \frac{\pi (2 n_1 + n_3)}{d_1}\,\,\, ;
\ee
\item $d_2 \neq 0$, $\left| \frac{d_1}{d_2} \right| \geq 1$: \\
inside the integration region in $\phi$ no pole may occur and the result
is
\bea
I &=&
- 2\, \pi \left[ \frac{n_2}{d_2} +\frac{n_4}{2\,d_2} +\frac{d_1\,n_3}{d_2^2} +
\frac{d_1^2\,n_4}{d_2^3} \right. \nonumber \\
&-& \left. \sqrt{\frac{d_1+d_2}{d_1-d_2}}\frac{1}{d_1+d_2} \left( n_1 +
\frac{d_1\,n_2}{d_2} + \frac{d_1^2\,n_3}{d_2^2} + \frac{d_1^3\,
n_4}{d_2^3} \right) \right]\,\,\, ;
\eea
\item $d_2 \neq 0$, $\left| \frac{d_1}{d_2} \right| < 1$: \\
in this case the integrand function may develop a pole and the integral
$I$ has to be evaluated in the principal value sense:
\be
I = \frac{-\pi\, \left[ 2\,d_1\, d_2\, n_3 + 2\,d_1^2\, n_4 + d_2^2\,
\left( 2\, n_2 + n_4 \right)\right]}{d_2 ^3}\,\,\, .
\ee
\end{enumerate}

The final expressions of the functions $I_{F_{1,2}}$ are reported in
Appendix \ref{vertapp}. Thus, the thermal vertex renormalization
correction to neutrino energy loss rate due to pair annihilations reads
\be
\Delta Q^V_{e^+e^-} = \Delta Q^{V,B}_{e^+e^-} \, + \, \Delta Q^{V,F}_{e^+
e^-}\,\,\, .
\label{dqvertex}
\ee

\subsubsection{Bremsstrahlung: $\gamma$ emission/absorption}
\label{bremcor}

As stated above, in order to eliminate the infrared divergencies present
in the radiative diagrams of Figures \ref{pair}b--\ref{pair}d it is
necessary to include the rates of processes where a photon is either
absorbed or emitted (see Figures \ref{pair}e--\ref{pair}h). The energy
loss rates due to $\gamma$ emission (absorption) $\Delta
Q^{E(A)}_{e^+e^-}$, are given by the sum of squared amplitudes of the
processes of Figures \ref{pair}e--\ref{pair}h, namely
\bea
\Delta Q^{E}_{e^+e^-} &=& \frac{1}{(2\pi)^9} \int \frac{d^3{\bvec{p}}_1}{
2E_1} \, \int\frac{d^3 \bvec{p}_2}{2E_2} \, \int\frac{d^3 \bvec{k}}{2
|{\bvec{k}}|} \, (E_1+E_2-|{\bvec{k}}|) F_-(E_1) \, F_+(E_2)\, [1 +
B(|{\bvec{k}}|)] \nonumber \\
&\times& \left\{ \frac{1}{(2 \pi)^2}\int \frac{d^3{\bvec{q}}_1}{ 2
\omega_1} \int \frac{d^3{\bvec{q}}_2}{2 \omega_2}\, \delta^4(p_1 + p_2 -
q_1 - q_2 - k)\, \sum_{{\mathrm spin,\alpha}} |M_{e^+ e^-
\rightarrow \nu_\alpha \ov{\nu}_\alpha \gamma}|^2 \right\}, \nonumber \\
\label{PE} \\
\Delta Q^{A}_{e^+e^-} &=& \frac{1}{(2\pi)^9} \int \frac{d^3 \bvec {p}_1}{
2E_1} \, \int\frac{d^3 \bvec{p}_2}{2E_2} \, \int\frac{d^3 \bvec{k}}{2
|{\bvec{k}}|}\, (E_1+E_2+|{\bvec{k}}|) F_-(E_1) \, F_+(E_2) \, B(|{\bvec{k}}|)
\nonumber \\
&\times& \left\{ \frac{1}{(2 \pi)^2}\int \frac{d^3{\bvec{q}}_1}{ 2
\omega_1} \int \frac{d^3{\bvec{q}}_2}{2 \omega_2}\,
\delta^4(p_1+p_2-q_1-q_2+k)\, \sum_{{\mathrm spin,\alpha}} |M_{e^+ e^-
\gamma \rightarrow \nu_\alpha \ov{\nu}_\alpha}|^2 \right\}. \nonumber \\
\label{PA}
\eea
Note that in the computation of the amplitudes for both processes, only
the vacuum part of the electron or positron propagator should enter since,
otherwise, we would have an electromagnetic vertex  with a real photon and
two real electrons or positrons which is not allowed by 4-momentum
conservation.

The term proportional to unity in the statistical factor
$(1+B(|{\bvec{k}}|))$ entering in Eq.(\ref{PE}) is responsible for the
spontaneous emission in vacuum. For small $|{\bvec{k}}|$ this contribution
has been already discussed in Section \ref{vacuum} (Soft--Bremsstrahlung),
hence we have to consider only the hard-photon emission part corresponding
to photon energies above the introduced cutoff $\epsilon$.

For the $\gamma$ emission, the term in curly brackets in Eq.(\ref{PE}) is
given by
\bea
\Phi^E &=& \frac{\alpha G_F^2}{24} \, \Theta (E_1 + E_2 - |{\bvec{k}}|) \,
\Theta (m_e^2 + p_1 \cdot p_2 - k \cdot (p_1+p_2)) \nonumber \\
&\times&\, \left[ \frac{\Phi^E_{1}}{(p_1 \cdot k)^2} \, + \, \frac{\Phi^E_{2}}
{(p_2 \cdot k)^2} \, + \, \frac{\Phi^E_{12}}{(p_1 \cdot k)(p_2 \cdot k)}
\right] \,\,\, ,
\label{tem}
\eea
while for $\gamma$ absorption we have
\bea
\Phi^A &=& \frac{\alpha G_F^2}{24} \, \left[ \frac{\Phi^A_{1}}{(p_1 \cdot
k)^2} + \frac{\Phi^A_{2}}{(p_2 \cdot k)^2} + \frac{\Phi^A_{12}}{(p_1 \cdot
k)(p_2 \cdot k)} \right]\,\,\, .
\label{tam}
\eea
The functions $\Phi^E_{1}$, $\Phi^E_{2}$, $\Phi^E_{12}$ are reported in
Appendix \ref{bremapp} while $\Phi^A_{1}$, $\Phi^A_{2}$, $\Phi^A_{12}$ for
the photon absorption are obtained from the corresponding $\Phi_i^E$ by
simply replacing the 4-momentum $k$ with $-k$. The $\Theta$-functions
appearing in Eq.(\ref{tem}), coming from the Lenard formula used in
performing neutrino momentum integration, give the kinematical conditions
for the process to occur. Obviously, for the $\gamma$ absorption we have
no kinematical constraints and we have omitted the superfluous
$\Theta$-functions \footnote{In this case the arguments of the
$\Theta$-functions are always positive for all values of the particle
4-momenta, as it can be easily checked, for example, in the
electron-positron center of mass frame.}.

In the limit $|{\bvec{k}}| \rightarrow 0$ the integrand function in
Eq.(\ref{PE}) (where the expression in Eq.(\ref{tem}) has been
substituted) shows some divergences which cancel the corresponding
infrared singularities in the radiative corrections considered previously.
To pick up the divergent terms, we note that the relevant expression
appearing in $\Delta Q^{E}_{e^+e^-}$ is the following:
\be
\frac{|{\bvec{k}}|^2}{|{\bvec{k}}|} \, \left( E_1 + E_2 - |{\bvec{k}}|
\right)\, \Phi^E\, [ 1 + B(|{\bvec{k}}|)]\,\,\,.
\ee
Thus, by expanding $\Phi^E$ in powers of $|{\bvec{k}}|$, the infrared-safe
part is obtained by subtracting the terms up to order 0 in $|{\bvec{k}}|$
to the expression proportional to the Bose function. As far as the factor
which does not contain $B(|{\bvec{k}}|)$ is concerned, we observe that no
change has to be performed because we must integrate this term for
$|{\bvec{k}}|$ ranging from the cutoff $\epsilon$ to the infinity, since
the soft part of the (vacuum) bremsstrahlung has been already considered
in Section \ref{vacuum}.

Analogous divergent terms for $|{\bvec{k}}| \rightarrow 0$ appear in the
integrand function for the energy loss induced by photon absorption. In
this case the relevant expression to be considered is
\be
\frac{|{\bvec{k}}|^2}{|{\bvec{k}}|} \, \left( E_1 + E_2 + |{\bvec{k}}|
\right)\, \Phi^A \,  B(|{\bvec{k}}|)\,\,\, ,
\ee
and, as in the previous case, the infrared-safe part is obtained by
subtracting terms up to order 0 in $|{\bvec{k}}|$.

Finally, the integrations in the electron, positron and photon 3-momenta
in Eq.(\ref{PE}) must be performed by taking into account the
$\Theta$-functions in Eq.(\ref{tem}). We choose the frame where the photon
momentum $\bvec{k}$ lies along the $z$-axis and ${\bvec{p}}_1$ is in the
$x$-$z$ plane. We also denote with $\theta_{1,2}$ the angle between
$\bvec{k}$ and ${\bvec{p}}_{1,2}$ and with $\phi$ the azimuthal angle of
the vector ${\bvec{p}}_2$.

The kinematical constraints can be implemented as follows. The integration
range for the variables $|{\bvec{p}_1}|$, $|{\bvec{p}_2}|$ is not limited
(they run over the entire $[0,\infty[$ interval), as well as those for
$\theta_1$, $\theta_2$ (ranging from $0$ to $\pi$). The integration in the
modulus of the photon momentum $|{\bvec{k}}|$ is bound to the region
$|{\bvec{k}}| \leq  E_1 + E_2$, thus implementing the condition coming
from the first $\Theta$-function in Eq.(\ref{tem}). The remaining
kinematical constraint can be written as:
\be
a_1\, \cos\phi\; \leq\; a_2\,\,\, ,
\label{phiext}
\ee
with ($s_{1,2}=\sin\theta_{1,2}$, $x_{1,2}=\cos\theta_{1,2}$)
\bea
a_1 &=& |{\bvec{p}_1}| |{\bvec{p}_2}| s_1 s_2 \,\,\, , \\
a_2 &=& E_1 E_2 - |{\bvec{p}_1}| |{\bvec{p}_2}| x_1 x_2 - |{\bvec{k}}|
(E_1 + E_2 - |{\bvec{p}_1}| x_1 - |{\bvec{p}_2}| x_2) + m_e^2\,\,\, .
\eea
This gives no effective condition when $a_1 =0$, $a_2\geq 0$ or $a_1 \neq
0$, $a_2/a_1 \geq 1$, thus leaving the integration over $\phi$ ranging
from $0$ to $2 \pi$, while Eq.(\ref{phiext}) cannot be fulfilled when
$a_1=0$, $a_2<0$ or $a_1\neq 0$, $a_2/a_1\leq -1$, thus resulting in
$\Delta Q^{E}_{e^+e^-}=0$ for these cases. In the remaining cases, that is
$a_1 \neq 0$, $-1 < a_2/a_1 < 1$, the integration interval in $\phi$ is
$[\alpha, 2 \pi - \alpha]$ with $\alpha = \arccos(a_2/a_1)$.

For the photon absorption case, since there are no kinematical
constraints, the integration on $|{\bvec{k}}|$ as well as on the angle
$\phi$ is not restricted to a given region but ranges over all possible
values.

The final expression for the sum of the neutrino energy loss rate induced
by photon emission and absorption is the following:
\bea
\Delta Q^{E}_{e^+e^-} + \Delta Q^{A}_{e^+e^-} =\frac{\alpha G_F^2}{3072\,
\pi^7} \, \int d |{\bvec{p}_1}| d |{\bvec{p}_2}| d |{\bvec{k}}| d x_1 d
x_2 \, \frac{|{\bvec{p}_1}|^2 |{\bvec{p}_2}|^2}{E_1 E_2} F_-(E_1) F_+(E_2)
\nonumber \\
\times \left[ I_{e0} \, \Theta (E_1 + E_2 -|{\bvec{k}}|) + (I_e\, \Theta
(E_1 + E_2 -|{\bvec{k}}|) + I_a) B(|{\bvec{k}}|) \right]\,\,\, ,
\label{qeqa}
\eea
where the function $I_{e0}, I_e, I_a$ are reported in Appendix
\ref{bremapp}.

\section{Neutrino photoproduction}
\label{photopro}

\begin{figure}
\begin{center}
\epsfysize=3.5cm
\epsfxsize=10.5cm
\epsffile{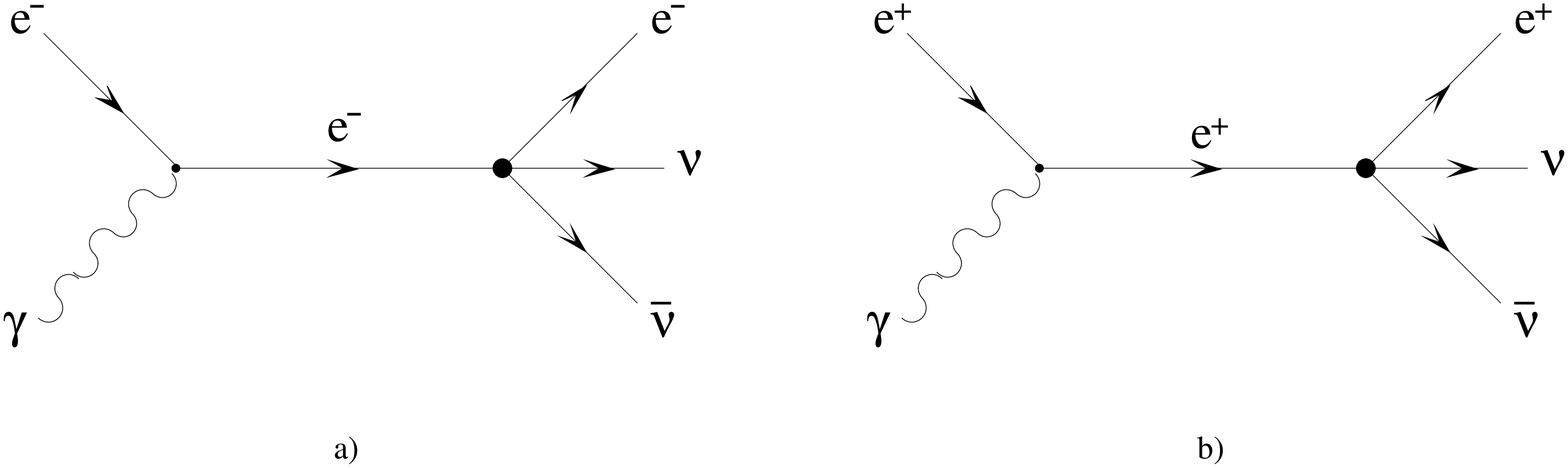}
\caption{Feynman diagrams for the neutrino photoproduction processes.}
\label{phot}
\end{center}
\end{figure}
Neutrino photoproduction processes $\gamma(k) \, + \, e^{\pm}(p_1)
\rightarrow e^{\pm}(p_2) \, + \, \nu_\alpha(q_1) \, + \,
\ov{\nu}_\alpha(q_2)$ contribute to the neutrino energy loss rate, at the
same perturbative order as the electromagnetic corrections to the pair
annihilation process. Thus, they must be included in a comprehensive study
of the star cooling rate. The relevant diagrams for photoproduction on
$e^-$ or $e^+$, shown in Figure \ref{phot}, are simply obtained by
crossing the corresponding ones for photon absorption in Figures
\ref{pair}g, \ref{pair}h, and the neutrino energy loss rates take the
following forms
\bea
Q_{\gamma e^-} &=& \frac{1}{(2\pi)^9} \int \frac{d^3{\bvec{p}}_1}{ 2E_1}\,
\int\frac{d^3 \bvec{p}_2}{2E_2} \, \int\frac{d^3 \bvec{k}}{2 E_\gamma}\,
(E_1-E_2+E_\gamma) F_-(E_1) \, [1- F_-(E_2)] \, B(E_\gamma) \nonumber \\
&\times& \left\{ \frac{1}{(2 \pi)^2}\int \frac{d^3{\bvec{q}}_1}{ 2
\omega_1} \int \frac{d^3{\bvec{q}}_2}{2 \omega_2}\, \delta^4(k + p_1 - p_2
- q_1 - q_2)\, \sum_{{\mathrm spin,\alpha}} |M_{\gamma e^- \rightarrow e^-
\nu_\alpha \ov{\nu}_\alpha }|^2 \right\}\,\,\, ,\nonumber \\
\label{phm} \\
Q_{\gamma e^+} &=& \frac{1}{(2\pi)^9} \int \frac{d^3 \bvec{p}_1}{ 2E_1}\,
\int\frac{d^3 \bvec{p}_2}{2E_2}\, \int\frac{d^3 \bvec{k}}{2 E_\gamma}\,
(E_2-E_1+E_\gamma) F_+(E_2)\, [1 -F_+(E_1)] \, B(E_\gamma) \nonumber \\
&\times& \left\{ \frac{1}{(2 \pi)^2}\int \frac{d^3{\bvec{q}}_1}{ 2
\omega_1} \int \frac{d^3{\bvec{q}}_2}{2 \omega_2}\, \delta^4(k + p_2 - p_1
- q_1 - q_2)\, \sum_{{\mathrm spin,\alpha}}|M_{\gamma e^+ \rightarrow e^+
\nu_\alpha \ov{\nu}_\alpha}|^2 \right\}\,\,\,. \nonumber \\
\label{php}
\eea
Differently from what happens for the corrections to the pair process,
analyzed in the previous Section, in the calculations of photoproduction
rates we must take into account the photon effective mass $m_\gamma$
depending on the density and temperature of the plasma. This effect could
be neglected in all the $e^+ e^- \rightarrow \nu\, \ov{\nu}$ radiative
corrections because it yielded an irrelevant change in the neutrino energy
loss rate, but it is important for the photoproduction processes which
exhibit a strong dependence on the background density in the high density
region.

In the following we consider a {\it massive} photon with the following
renormalized mass \cite{masood}
\be
m_\gamma^2  \; = \;
\frac{4 \alpha}{\pi}\, \int_0^\infty \, d|{\bvec{k}}|\,
\frac{|{\bvec{k}}|^2}{E_k} \, \left( F_+(E_k) + F_-(E_k)
\right)\,\,\, ,
\label{mgam}
\ee
where we have adopted the same notations of Eq.(\ref{dme}) and the
completeness relation
\be
\sum_\lambda \epsilon_\mu^{(\lambda)} \epsilon_\nu^{(\lambda)} = -
g_{\mu \nu} \, + \, \frac{k_\mu k_\nu}{k^2}\,\,\,.
\ee
Note that a finite photon mass eliminates infrared divergencies due to the
Bose distribution function in the rates. For the sake of brevity, in the
following we describe the calculations for photoproduction on electron
only. The corresponding ones for photoproduction on positron can be
obtained by replacing $\xi_e \rightarrow -\xi_e$.

The term in curly brackets in Eq.(\ref{phm}) is given by
\bea
\Phi_{\mathrm ph} &=& \frac{8\, \alpha\, G_F^2}{3\, m_\gamma^2}~ \Theta (E_1
- E_2 + E_\gamma)\, \Theta (2\, m_e^2 + m_\gamma^2 - 2\, p_1 \cdot p_2 + 2\,
k \cdot (p_1-p_2)) \nonumber \\
&\times& \left[ \frac{\Phi_{\mathrm ph}^{1}}{(m_\gamma^2 + 2\, p_1 \cdot k)^2}\,
+\, \frac{\Phi_{\mathrm ph}^{2}}{(m_\gamma^2 - 2\, p_2 \cdot k)^2}\, +\, \frac{2\,
\Phi_{\mathrm ph}^{12}}{(m_\gamma^2 + 2\, p_1 \cdot k)(m_\gamma^2 - 2\, p_2
\cdot k)} \right]\,\,\, .
\label{tmph}
\eea
The squared amplitude function $\Phi_{\mathrm ph}^1$, $\Phi_{\mathrm
ph}^2$, $\Phi_{\mathrm ph}^{12}$, are reported in Appendix \ref{photoapp}.
The $\Theta$-functions appearing in Eq.(\ref{tmph}), coming from the
Lenard formula used in performing neutrino momentum integration, give the
kinematical conditions for the process to occur. The remaining
integrations on the incoming electron and photon and outgoing electron
3-momenta in Eq.(\ref{phm}) are performed, in close analogy to what done
in Section \ref{bremcor}, by choosing the frame where the photon momentum
$\bvec{k}$ lies along the $z$-axis and ${\bvec{p}}_1$ is in the $x$-$z$
plane (denoting with $\theta_{1,2}$ the angle between $\bvec{k}$ and
${\bvec{p}}_{1,2}$ and with $\phi$ the azimuthal angle of the vector
${\bvec{p}}_2$). In particular the integration in $|{\bvec{p}_1}|$,
$|{\bvec{k}}|$, $\theta_1$, $\theta_2$ ranges over all possible values for
these variables, while the integration field for $|{\bvec{p}_2}|$ is
limited by the first $\Theta$-function in Eq.(\ref{tmph}):
\be
0 \; \leq \; |{\bvec{p}_2}| \; \leq \; \sqrt{|{\bvec{p}_1}|^2 + E_\gamma^2
+ 2 E_\gamma E_1} \equiv p_2^{\,\mathrm max}\,\,\,.
\ee
The remaining kinematical constraint, which has to be used to perform the
integration on the angle $\phi$, can be written as:
\be
b_1 \, \cos \phi \; \geq \; b_2\,\,\, ,
\label{conphi2}
\ee
with ($s_{1,2}=\sin\theta_{1,2}$, $x_{1,2}=\cos\theta_{1,2}$)
\bea
b_1 &=& 2\, |{\bvec{p}_1}|\, |{\bvec{p}_2}|\, s_1\, s_2\,\,\, , \\
b_2 &=& 2\, E_1\, E_2 - 2\, |{\bvec{p}_1}|\, |{\bvec{p}_2}|\, x_1\, x_2 -
2\, E_\gamma\, (E_1 - E_2) + \nonumber \\
&~& + 2\, |{\bvec{k}}|\, (|{\bvec{p}_1}|\, x_1 - |{\bvec{p}_2}|\, x_2) - 2\,
m_e^2 - m_\gamma^2\,\,\, .
\eea
Note that the condition (\ref{conphi2}) is not fulfilled when $b_1=0$,
$b_2>0$ or $b_1 \neq 0$, $b_2/b_1 \geq 1$, thus resulting in $Q_{\gamma
e^-}=0$, while it is trivially realized for $b_1 =0$, $b_2 \leq 0$, where
case the integration field for $\phi$ is $[0,2\pi]$. The constraint
(\ref{conphi2}) can be non trivially satisfied only for $b_1 \neq 0$, $-1
< b_2/b_1 <1$, when the integration region for $\phi$ is $[0,\beta] \cup
[2\pi -
\beta,2\pi]$ with $\beta = \arccos(b_2/b_1)$. The final expression
for the energy loss rate induced by neutrino photoproduction on
electron is the following
\bea
Q_{\gamma e^-} &=& \frac{\alpha\, G_F^2}{3072\, \pi^7} \, \int_0^\infty d
|{\bvec{p}_1}| d |{\bvec{k}}| \int_0^{p_2^{\,\mathrm max}} d|{\bvec{p}_2}|
\int_{-1}^{1} d x_1 d x_2\, \frac{|{\bvec{p}_1}|^2 |{\bvec{p}_2}|^2}{E_1 E_2}\,
\nonumber \\
&\times& F_-(E_1)\, [1 - F_-(E_2)]\, B(E_\gamma)\, I_{\mathrm ph}\,\,\, ,
\label{qphom}
\eea
where the function $I_{\mathrm ph}$ is reported in Appendix \ref{photoapp}.

The corresponding expression for the rate induced by neutrino
photoproduction on posi\-tron is easily obtained from Eq.(\ref{qphom})
with simple substitutions,
\bea
Q_{\gamma e^+} &=& \frac{\alpha\, G_F^2}{3072\, \pi^7} \, \int_0^\infty d
|{\bvec{p}_2}| d |{\bvec{k}}| \int_0^{p_1^{\,\mathrm max}} d|{\bvec{p}_1}|
\int_{-1}^{1} d x_1 d x_2\, \frac{|{\bvec{p}_1}|^2 |{\bvec{p}_2}|^2}{E_1 E_2}\,
\nonumber \\
&\times& F_+(E_2)\, [1 - F_+(E_1)]\, B(E_\gamma)\, I_{\mathrm
ph}^\prime\,\,\, ,
\label{qphop}
\eea
and $I_{\mathrm ph}^\prime$ is obtained from $I_{\mathrm ph}$ with $p_1
\leftrightarrow p_2$\footnote{Note that the kinematics of the process is
the same as for the photoproduction on electron by replacing the 4-vector
$p_!$ with $p_2$.}. Thus by using the expressions (\ref{qphom}) and
(\ref{qphop}), one gets the total energy loss rate due to neutrino
photoproduction as
\be
Q_{\gamma e} = Q_{\gamma e^-} + Q_{\gamma e^+}\,\,\, .
\label{photop}
\ee

\section{Plasmon decay}
\label{plasmpro}

\begin{figure}
\begin{center}
\epsfysize=3.5cm
\epsfxsize=6.3truecm
\epsffile{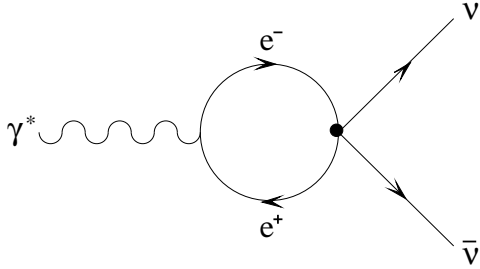}
\caption{Feynman diagram for the plasmon decay into a neutrino-antineutrino pair.}
\label{plasm}
\end{center}
\end{figure}
The plasma process $\gamma^\ast(k) \rightarrow \nu_\alpha(q_1) \, + \,
\ov{\nu}_\alpha(q_2)$ is of higher order in $\alpha$ with respect to the
pair annihilation process (it is of order $\alpha \sqrt{\alpha} G_F^2$)
but, nevertheless, its contribution to neutrino energy loss rate is
dominant for a wide region of stellar temperatures and densities
\cite{Raffelt95}. Then, in this Section, we discuss this process and
evaluate the corresponding energy loss rate.

The decay $\gamma^\ast \rightarrow \nu_\alpha \, + \, \ov{\nu}_\alpha$
takes place only if the photon momentum is a time-like 4-vector such that
$E_\gamma > |{\bvec{k}}|$. To see when this constraint applies, we have
first to calculate the dispersion relation for a photon in a thermal
plasma and this is recalled in Appendix \ref{plasmapp}. For transverse (T)
propagation modes there is no effective constraint, while for longitudinal
(L) ones the decay occurs only for $|{\bvec{k}}| < |{\bvec{k}}|_{\rm
max}$, where $|{\bvec{k}}|_{\rm max}$ is the maximum plasmon momentum in
the bath whose expression is reported in Appendix \ref{plasmapp}.

At lowest order the process $\gamma^\ast \rightarrow \nu_\alpha \, + \,
\ov{\nu}_\alpha$ is described by the Feynman diagram in Fig. \ref{plasm}
and the loss rate is given by
\be
Q_{\gamma^\ast} = Q^L_{\gamma^\ast} \, + \, Q^T_{\gamma^\ast}\,\,\, ,
\label{qpltot}
\ee
where the partial rates
\bea
Q^{L,T}_{\gamma^\ast} &=& \frac{1}{(2\pi)^3} \int\frac{d^3 \bvec{k}}{2
E_{\gamma L,T}} \, E_{\gamma L,T} \, B(E_{\gamma L,T}) \nonumber \\
&\times& \left\{ \frac{1}{(2 \pi)^2}\int \frac{d^3{\bvec{q}}_1}{2
\omega_1} \int \frac{d^3{\bvec{q}}_2}{2 \omega_2} \, \delta^4(k-q_1-q_2)\,
\sum_{\alpha} |M_{\gamma^\ast \rightarrow \nu_\alpha
\ov{\nu}_\alpha}^{L,T}|^2 \right\}\,\,\, ,
\label{pllt}
\eea
correspond to longitudinal and transverse photon mode propagation,
respectively. The computation of the invariant amplitude for the
considered decay can proceed through an effective  photon-neutrino
interaction
\be
M_{\gamma^\ast \rightarrow \nu \ov{\nu}}^\lambda = \ov{u}(q_1) \gamma^\mu
(1-\gamma_5) v(q_2) \, \Gamma_{\alpha \mu}(k)\,
\epsilon^\alpha_\lambda(k)\,\,\, ,
\label{mpl}
\ee
where $\epsilon^\alpha_\lambda(k)$ is the polarization 4-vector for
longitudinal ($\lambda = L$) or transverse ($\lambda = T$) thermal photons
and, by considering only the thermal on-shell propagation of $e^{\pm}$ in
the loop in Figure \ref{plasm}. Thus one gets
\be
\Gamma_{\alpha \mu}(k) = i \, \frac{e G_F}{\sqrt{2}} \int \frac{d^4p}{(2
\pi)^4} \left[ \cv' S_{\alpha \mu} - i \ca' A_{\alpha \mu} \right] \left[
\frac{\Gamma_F(p-k)}{p^2 - m_e^2} + \frac{\Gamma_F(p)}{(p-k)^2 - m_e^2}
\right]
\label{gampl}
\ee
\bea
S_{\alpha \mu} &=& (p-k)_\alpha p_\mu + (p-k)_\mu p_\alpha + p \cdot k
g_{\alpha
\mu}\,\,\, , \\
A_{\alpha \mu} &=& \epsilon_{\alpha \mu \rho \sigma} k^\rho p^\sigma\,\,\,
.
\eea
The quantity in curly brackets in Eq.(\ref{pllt}) is thus given by
\cite{Braaten93}:
\bea
\Phi_{\gamma^\ast}^{L,T} &=& - \frac{(E_{\gamma L,T})^2 - k^2}{3 \pi} \,
g_{\mu \nu}\, P^{\mu \nu}_{L,T}\,\,\, , \\
P^{\mu \nu}_{L} &=& - \, Z_L \, Q_{\alpha \beta} \, \Gamma^{\alpha \mu}
\left( \Gamma^{\beta \nu} \right)^\ast \,\,\, , \\
P^{\mu \nu}_{T} &=& - 2 Z_T\, R_{\alpha \beta}\, \Gamma^{\alpha \mu}
\left( \Gamma^{\beta \nu} \right)^\ast\,\,\, , \\
Q^{\alpha \beta} &=& \frac{K^2}{K^2 - (u \cdot K)^2} \left( u^\alpha -
\frac{u \cdot K}{K^2} K^\alpha \right) \left( u^\beta - \frac{u \cdot
K}{K^2} K^\beta \right) \,\,\, , \label{qmunu} \\
R^{\alpha \beta} &=& g^{\alpha \beta} - \frac{K^\alpha K^\beta}{K^2} -
Q^{\alpha \beta} \,\,\, , \label{rmunu}
\eea
where $u^\alpha=(1,{\bvec 0})$ is the medium 4-velocity and we have used
the orthogonality condition $K_\alpha \Gamma^{\alpha \mu} =0$, with
$K=(E_\gamma,{\bf k})$ and $E_\gamma = \sqrt{|{\bvec{k}}|^2+ m_\gamma^2}$.
Note that for longitudinal and transverse photon modes in a plasma we have
\cite{Weldon,Braaten93}
\bea
\sum_{\lambda=0} \epsilon^\alpha_\lambda \ \epsilon^{\beta \ast}_\lambda &=&
- \ Q^{\alpha \beta}\, Z_L \,\,\, ,\\
\sum_{\lambda={\pm}1} \epsilon^\alpha_\lambda \ \epsilon^{\beta
\ast}_\lambda &=& - 2 R^{\alpha \beta}\, Z_T\,\,\, ,
\eea
respectively, where the $k-$dependent functions $Z_{L,T}$ can be found in
Appendix \ref{plasmapp}.  The energy loss rates are, then
\cite{Braaten93,Raffelt95}
\bea
Q^L_{\gamma^\ast} &=& \frac{G_F^2 \cv^{\prime 2}}{96 \pi^4 \alpha}\,
\int_0^{|{\bvec{k}}|_{\rm max}} d|{\bvec{k}}|\, |{\bvec{k}}|^2\,
B(E_{\gamma L})\, Z_L\, \Pi_L^3\,\,\, , \label{qpll} \\
Q^T_{\gamma^\ast} &=& \frac{G_F^2 \cv^{\prime 2}}{48 \pi^4 \alpha}\,
\int_0^{\infty} d|{\bvec{k}}|\, |{\bvec{k}}|^2\, B(E_{\gamma T})\, Z_T\,
\Pi_T^3\,\,\, , \label{qplt}
\eea
with $\Pi_{L,T}$ reported in Appendix \ref{plasmapp}.

\section{Results and conclusions}

\begin{figure}
\begin{center}
\epsfysize=8cm
\epsfxsize=12cm
\epsffile{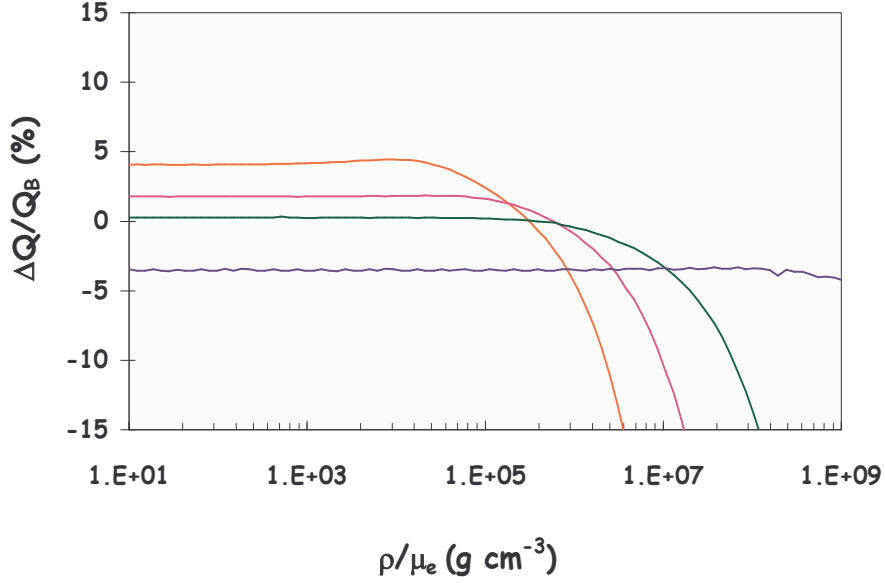}
\caption{The total radiative corrections normalized to the Born approximation
result for the pair annihilation process for $T=10^8,10^{8.5},10^9,10^{10}
~\kelvin$ (from top to bottom).}
\label{paircorr}
\end{center}
\end{figure}
In this paper we have presented an exhaustive computation of the energy
loss rates in neutrinos of the stellar interior. A consistent analysis of
all leptonic processes up to order $\alpha$ in the electromagnetic fine
structure constant has been performed: neutrino energy loss rate due to
pair annihilations, evaluated in Born approximation, $Q_{e^+e^-}^B$
(\ref{qpair}), has been corrected including both $vacuum$ and $thermal$
radiative corrections, the latter being computed in the real time
formalism. By using Eqs.(\ref{Qnue}), (\ref{qpair}), (\ref{dqt0}),
(\ref{dqmass}), (\ref{dqwave}), (\ref{dqvertex}), and (\ref{qeqa}), the
total radiative correction $\Delta Q_{e^+ e^-}$ results
\bea
\Delta Q_{e^+ e^-} &\equiv& Q_{e^+e^-} - Q_{e^+e^-}^B \nonumber \\
&& \!\!\!\!\!\!\!\!\!\!\!\!\!\!\!\!\!\! =
\Delta Q^{T=0}_{e^+e^-} + \Delta Q^M_{e^+e^-} + \Delta Q^W_{e^+e^-} +
\Delta Q^V_{e^+e^-} + \Delta Q^{E}_{e^+e^-} + \Delta Q^{A}_{e^+e^-}\,\,\,
.
\eea
In Figure \ref{paircorr}, we plot the ratio $\Delta Q_{e^+
e^-}/Q_{e^+e^-}^B$ as functions of the plasma density for some values of
the temperature, namely $T=10^8,10^{8.5},10^9,10^{10} ~\kelvin$. The
corrections are found to be of the order of few percent and negative for
high temperatures, implying that for these temperatures the energy loss is
sensibly decreased. For fixed temperature, $\Delta Q_{e^+
e^-}/Q_{e^+e^-}^B$ goes to a constant value for low density. This can be
easily understood, since in this limit the plasma is weakly degenerate,
and therefore the energy loss rate depends on temperature only. At large
densities the ratio $\Delta Q_{e^+ e^-}/Q_{e^+e^-}^B$ decreases and
reaches larger negative values. However, for such high densities the pair
annihilation rates are exceedingly small and thus this process gives only
a marginal contribution to the star cooling.

\begin{figure}
\begin{center}
\bt{ll}
\epsfxsize=7.5cm
\epsfysize=10cm
\epsffile{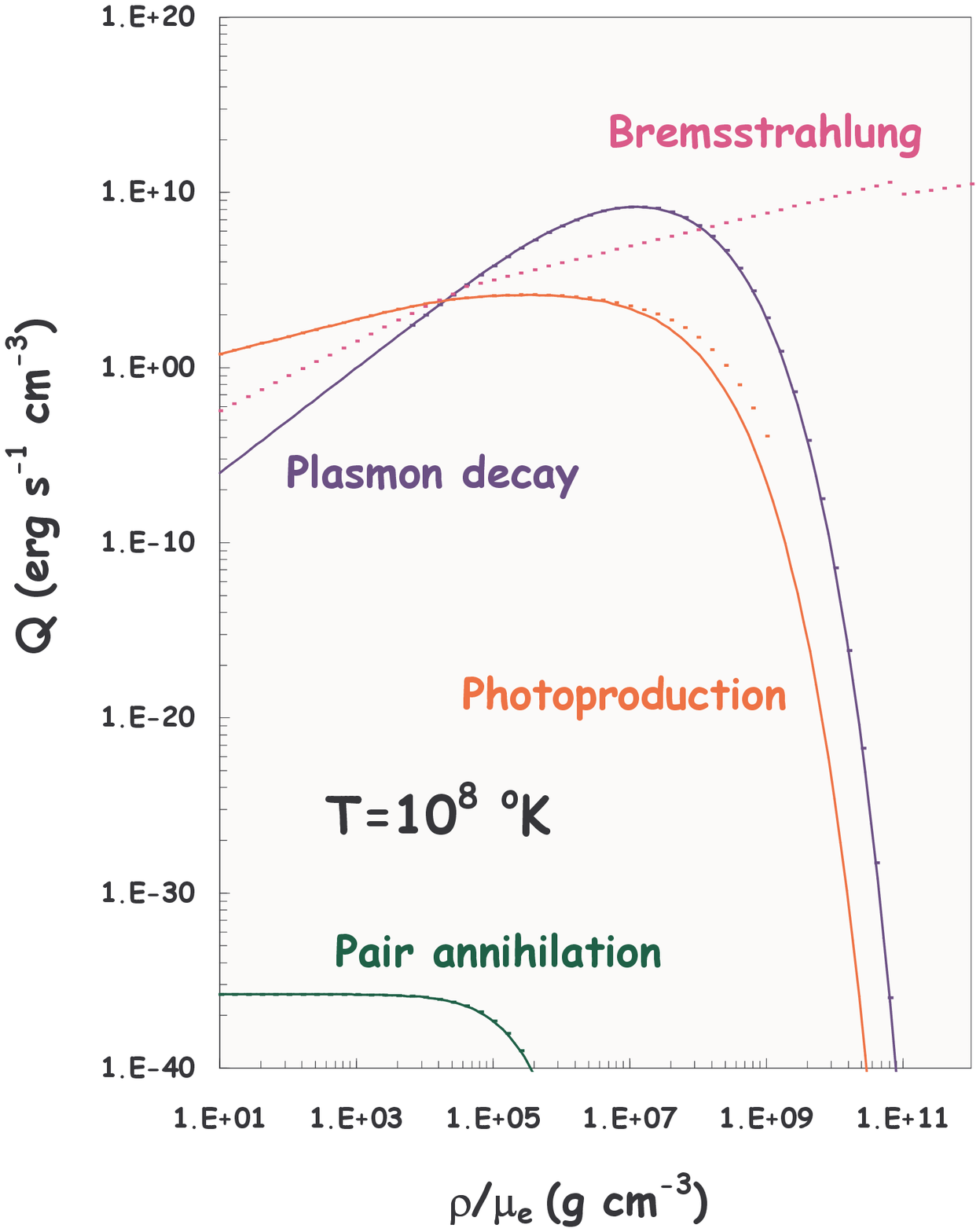}
&
\epsfxsize=7.5cm
\epsfysize=10cm
\epsffile{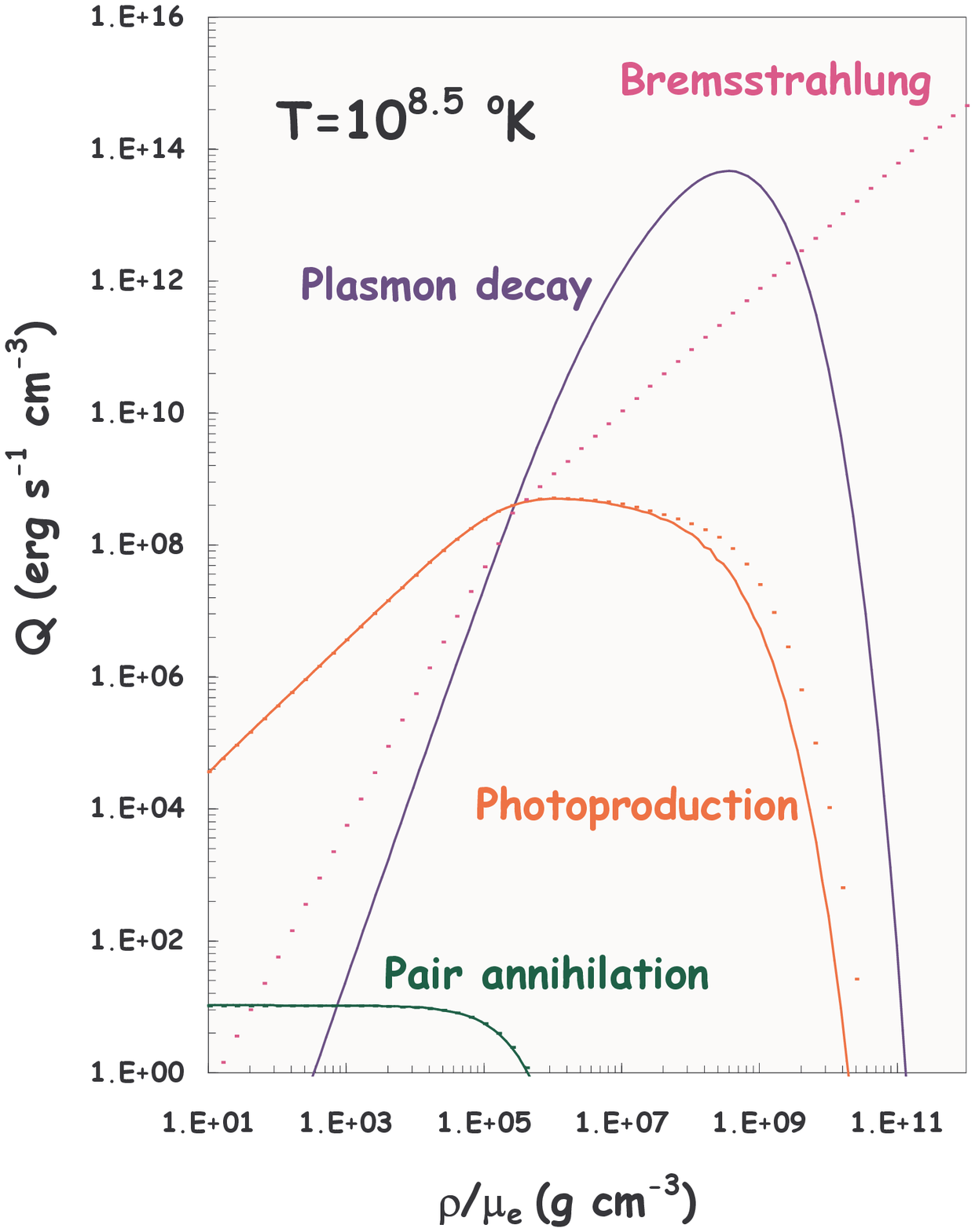}
\\
\epsfxsize=7.5cm
\epsfysize=10cm
\epsffile{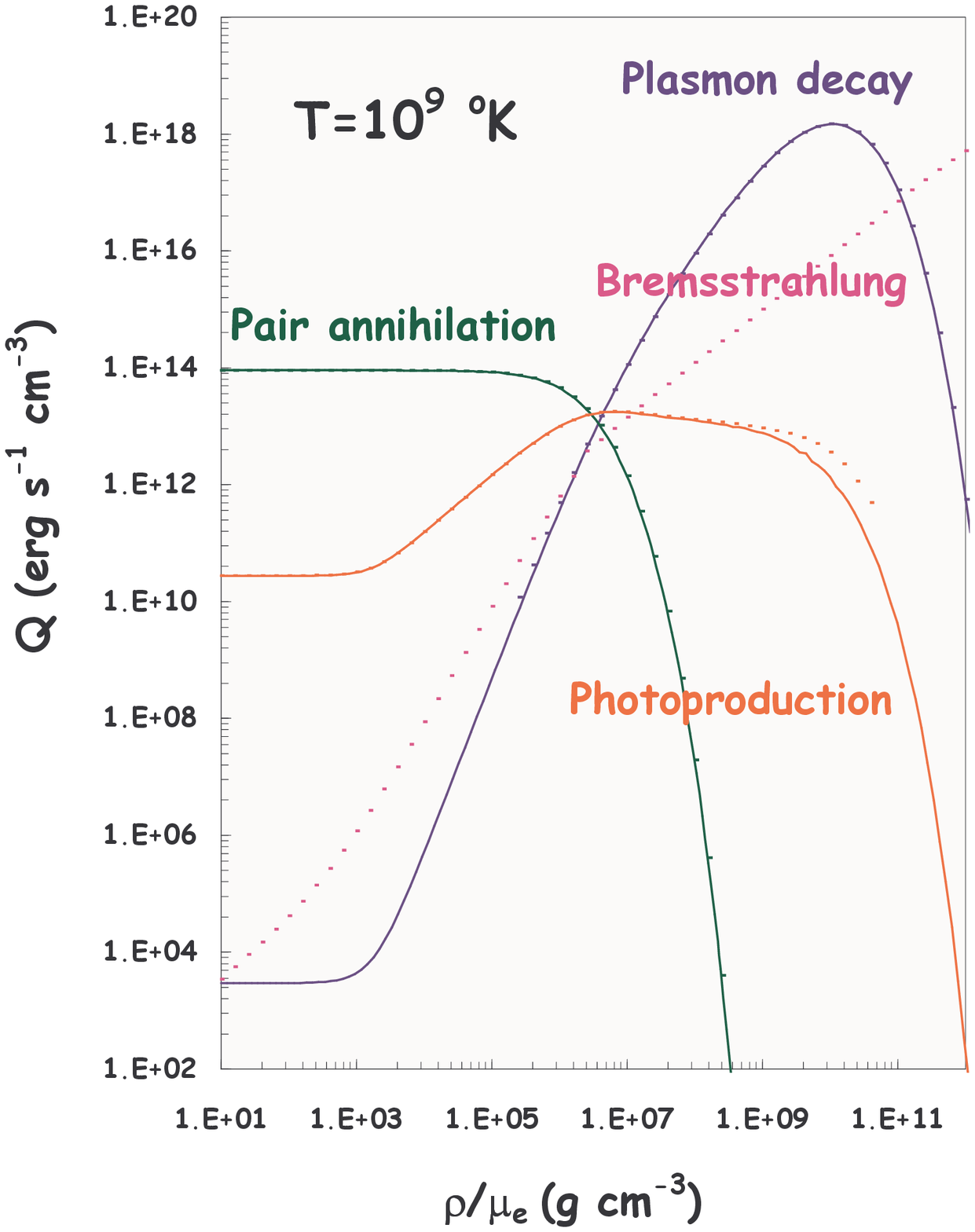}
&
\epsfxsize=7.5cm
\epsfysize=10cm
\epsffile{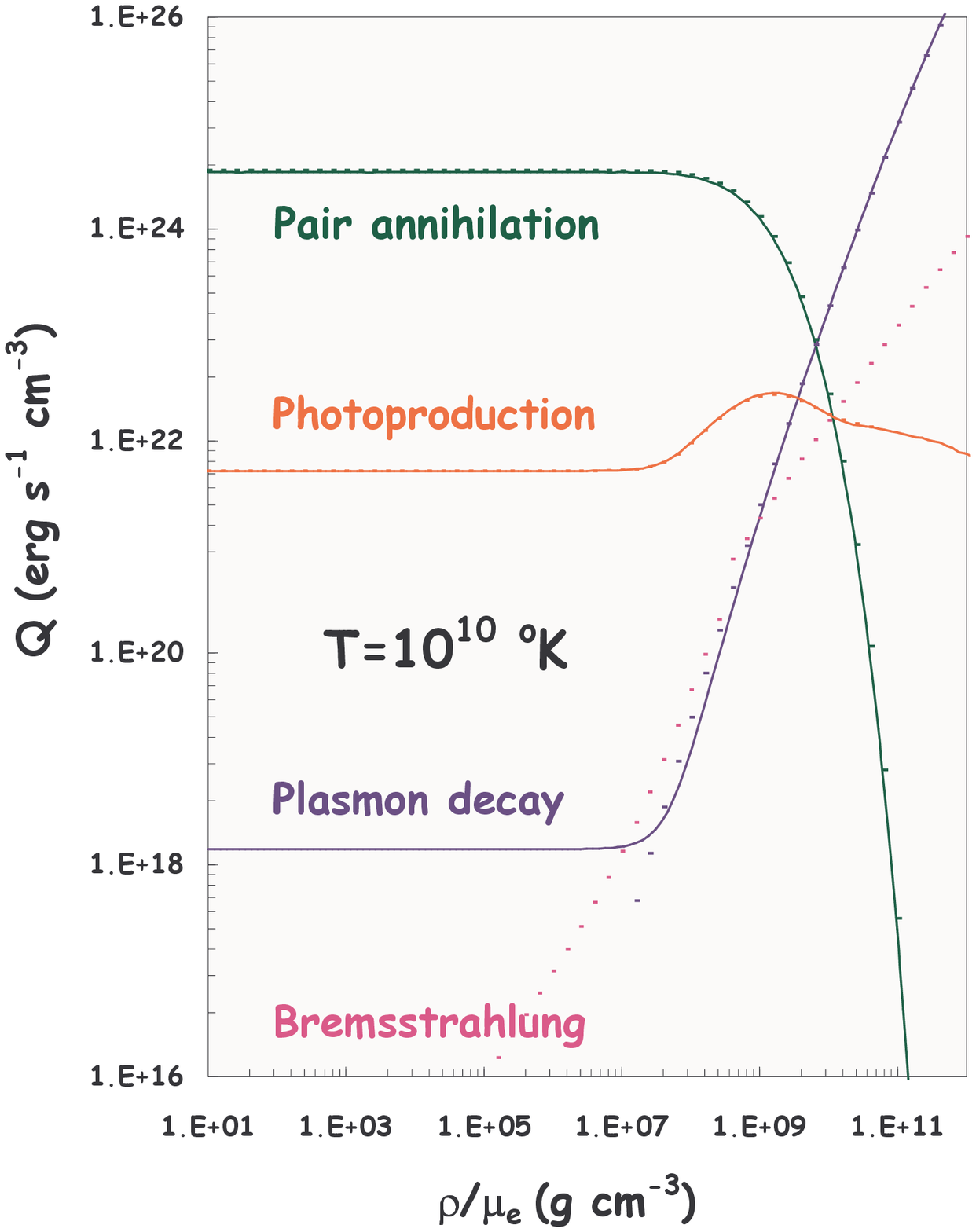}
\et
\end{center}
\caption{The energy loss rate versus $\rho/\mu_e$ due to pair
annihilation including radiative corrections $Q_{e^+ e^-}$ ,
photoproduction $Q_{\gamma e}$ and plasmon decay $Q_{\gamma^\ast}$ (solid
lines) for several temperatures (see for definitions Eqs.(\ref{Qnue}),
(\ref{photop}), (\ref{qpltot})). The dotted lines refer to the analogous
results of Ref. \cite{Itoh96}, where the rate for bremsstrahlung on nuclei
is also computed. The effect of $\Delta Q_{e^+ e^-}$ to pair annihilation
cannot be appreciated in this logarithmic scale but for the largest
temperature $10^{10} \, \kelvin$.}
\label{qcomp}
\end{figure}
In order to perform a comprehensive study, we have also recalculated the
contribution from $\nu$--photoproduction and plasmon decay processes. The
corresponding numerical results are presented in Figure \ref{qcomp}, where
we show the several contributions to total neutrino energy loss rate. When
possible, we also show for comparison the results of Ref. \cite{Itoh96}.
In particular we find a fair agreement with the results reported in
literature. For completeness we also show the energy loss rate due to
bremsstrahlung on nuclei, denoted with $Q_{eZ}$, which have been produced
using the analytic fitting formula of Ref. \cite{Itoh96}.

In summary, Figure \ref{qcomp} shows that, as well known, for large
temperatures and not too high densities, pair annihilation dominates over
the other three processes, while for low densities $\nu$-photoproduction
dominates over plasmon decay and bremsstrahlung on nuclei. On the other
hand, for large densities the most relevant process is plasmon decay,
whose rate however, along with those of all other processes, rapidly falls
down for extremely high densities. This is a genuine plasma effect as
noted in \cite{EMMPP}. Consider, for example, the behaviour of
$\nu$-photoproduction energy loss. As already noted in \cite{Beaudet67b},
the decrease for very large densities is achieved only if one consistently
takes into account the increasingly large photon thermal mass. In fact
with a massless photon the $\nu$-photoproduction curves in Figure
\ref{qcomp} would rather reach a constant value. The main effect of
$m_\gamma^2$ is a lowering of the values of the Bose distribution function
for photons, i.e. a smaller number of thermal photons. This reduces the
energy loss rate induced by $\nu$-photoproduction.

\begin{figure}
\begin{center}
\epsfysize=8.5cm
\epsfxsize=8.5cm
\epsffile{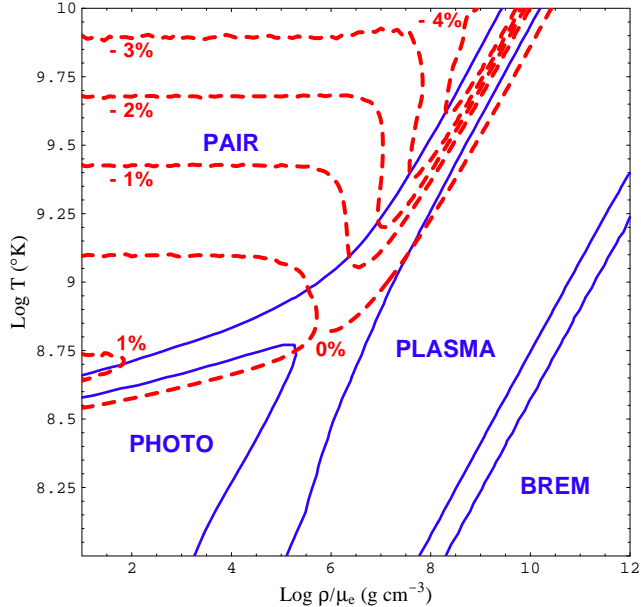}
\caption{The regions in the $T-\rho/\mu_e$ plane where each of the processes
i)-iv) contribute for more than $90 \%$ to the total energy loss rate. We
also show the contours for the relative correction $\Delta Q/Q^0_{Tot}$
(see text) for the values $1\%,0\%,-1\%,-2\%,-3\%,-4 \%$.}
\label{qcontour}
\end{center}
\end{figure}

In Figure \ref{qcontour} we show the regions in the temperature-density
plane where a given process contributes to the total energy loss rate
$Q_{Tot}= Q_{e^+ e^-} + Q_{\gamma e} + Q_{\gamma^\ast} + Q_{eZ}$
(including radiative corrections to pair annihilation) for more than 90\%.
We also summarize there our results on the radiative corrections to pair
annihilation processes, by plotting the contours corresponding to $\Delta
Q_{e^+ e^-}/Q_{Tot}^0=1\%,0\%,-1\%,-2\%,-3\%,-4 \%$, where $Q_{Tot}^0$ is
the total emission rate with pair annihilation calculated in Born
approximation. These contours lie almost entirely in the region where the
pair annihilation process gives the main contribution to the total energy
loss rate. This result may affect the late stages of evolution of very
massive stars by changing their configuration at the onset of Supernova
explosion.

\section*{Acknowledgements}

The authors would like to thank S. Chieffi, G. Imbriani, M. Limongi, M.
Passera, O. Straniero for useful discussions and comments.

\newpage

\appendix

\section{Vertex correction functions}
\label{vertapp}
\setcounter{equation}0

With the same notations of Section \ref{vertcor}, the function $I_B$
appearing in Eq.(\ref{dqbv}) is given by:
\bea
I_B &=& \frac 14 \int_0^{2 \pi} d \phi \, \left[ \left. \frac{\Phi_V^{IS}}
{\ p_1 \cdot k\, p_2 \cdot k} \right|_{k_0=|{\bvec{k}}|} +  \left.
\frac{\Phi_V^{IS}}{\ p_1 \cdot k \, p_2 \cdot k}
\right|_{k_0=-|{\bvec{k}}|} \right] \nonumber \\
&=& 256\,{{\ca}}^{\!\!\! \prime  2}\, \pi\, \left( 2\, E_1\, E_2 - 2\,
{|{\bvec{p}_1}|}\, {|{\bvec{p}_2}|}\, {x_1}\,{x_2} - m_2^2\, \frac{{E_1} -
{|{\bvec{p}_1}|}\,{x_1}}{E_2 - {|{\bvec{p}_2}|}\,{x_2}} - m_e^2\,
\frac{{E_2} - {|{\bvec{p}_2}|}\,{x_2}}{E_1 - {|{\bvec{p}_1}|}\,{x_1}}
\right) \nonumber \\
&& + 512\, {{\cv}}^{\!\!\! \prime  2}\, \pi\, \left( m_e^2 + E_1\, E_2
-{|{\bvec{p}_1}|}\, {|{\bvec{p}_2}|}\, x_1\, x_2 \right)\,\,\, ,
\eea
where $\Phi_V^{IS}$ is $\Phi_V$ once subtracted the terms of order 0 and 1
in $k$.

On the other side, after performing the integration involved in
Eq.(\ref{ii}), one obtains in the three cases of Section \ref{vertcor}:
\begin{enumerate}
\item $d_2=0$;
\bea
&I_{F_1}& = \frac{-128\, \pi}{\left(m_e^2-k\cdot p_1\right)\, \left[k \cdot (p_1 +
p_2) - E_1\, E_2-m_e^2+|{\bvec{p}_1}|\, |{\bvec{p}_2}|\, x_1\, x_2\right]}
\nonumber \\
&\times& \left\{ {{\cv}}^{\!\!\! \prime  2} \left[ |{\bvec{p}_1}|^2\,
|{\bvec{p}_2}|^2\, \left(-1 + {x_1}^2 \right)\, \left(-1 + x_2^2 \right)
\right. \right. \nonumber \\
&\times& \left(6\, E_1\, E_2 - 2\, k \cdot p_2 + 5\, m_e^2 - 6\, |{\bvec{p}_1}|\,
|{\bvec{p}_2}|\, x_1\, x_2 \right) \nonumber \\
&+&  2\, \left(E_1\, E_2 + m_e^2 - |{\bvec{p}_1}|\, |{\bvec{p}_2}|\, x_1\, x_2
\right)\, \left( 2\, E_1^2\, E_2^2 - 5\, k \cdot p_2\, m_e^2 \right. \nonumber \\
&+& k \cdot p_1\, \left(2\, k \cdot p_2 + 3\, m_e^2 \right) - 3\, m_e^4 - 3\, m_e^2\,
|{\bvec{p}_1}|\, |{\bvec{p}_2}|\, x_1\, x_2 \nonumber \\
&+& E_1\, E_2\, \left(-2\, k \cdot p_2 + 3\, m_e^2 - 4\, |{\bvec{p}_1}|\,
|{\bvec{p}_2}|\, x_1\, x_2 \right) \nonumber \\
&+& \left. \left. 2\, |{\bvec{p}_1}|\, |{\bvec{p}_2}|\, x_1\, x_2\, \left( k \cdot
p_2 + |{\bvec{p}_1}|\, |{\bvec{p}_2}|\, x_1\, x_2 \right) \right) \right] \nonumber \\
&+& {{\ca}}^{\!\!\! \prime  2} \left[ |{\bvec{p}_1}|^2\, |{\bvec{p}_2}|^2\,
\left(-1 + x_1^2 \right)\, \left(-1 + x_2^2 \right) \right. \nonumber \\
&\times& \left(6\, E_1\, E_2 - 2\, k \cdot p_2 + m_e^2 - 6\, |{\bvec{p}_1}|\,
|{\bvec{p}_2}|\, x_1\, x_2 \right) \nonumber \\
&+& 2\, \left(2\, E_1^3\, E_2^3 + 2\, k \cdot p_1\, m_e^4 - (k \cdot p_1)^2\,
m_e^4 - {k \cdot p_2}^2\, m_e^4 - m_e^6 \right. \nonumber \\
&-&2\, k \cdot p_1\, \left( k \cdot p_2 + m_e^2 \right)\, |{\bvec{p}_1}|\,
|{\bvec{p}_2}|\, x_1\, x_2 + 2\, m_e^4\, |{\bvec{p}_1}|\, |{\bvec{p}_2}|\,
x_1\, x_2 \nonumber \\
&+& E_1^2\, E_2^2\, \left(-2\, k \cdot p_2 + m_e^2 - 6\, |{\bvec{p}_1}|\,
|{\bvec{p}_2}|\, x_1\, x_2 \right) \nonumber \\
&+& |{\bvec{p}_1}|\, |{\bvec{p}_2}|\, x_1\, x_2\, \left( |{\bvec{p}_1}|\,
|{\bvec{p}_2}|\, x_1\, x_2\, \left( m_e^2 - 2\, |{\bvec{p}_1}|\,
|{\bvec{p}_2}|\, x_1\, x_2 \right) \right. \nonumber \\
&+& \left. k \cdot p_2\, \left(2\, m_e^2 - 2\, |{\bvec{p}_1}|\,
|{\bvec{p}_2}|\, x_1\, x_2 \right) \right) \nonumber \\
&+& 2\, E_1\, E_2\, \left( k \cdot p_1\, \left( k \cdot p_2 + m_e^2 \right) -
m_e^4 + k \cdot p_2\, \left(-m_e^2 + 2\, |{\bvec{p}_1}|\, |{\bvec{p}_2}|\,
x_1\, x_2 \right) \right. \nonumber \\
&+& \left. \left. \left. \left. |{\bvec{p}_1}|\, |{\bvec{p}_2}|\, x_1\, x_2\,
\left(-m_e^2 + 3\, |{\bvec{p}_1}|\, |{\bvec{p}_2}|\, x_1\, x_2 \right)
\right) \right) \right] \right\}
\eea
\item $d_2 \neq 0$, $\left| \frac{d_1}{d_2} \right| \geq 1$;
\bea
&I_{F_1}& = \frac{256\, \pi}{m_e^2-k\cdot p_1}\, \left\{ {{\cv}}^{\!\!\! \prime
2}\, \left[ -3\, m_e^4 + 2\, E_1^2\, E_2^2 + 2\, (k \cdot p_1)^2 - 4\,
m_e^2\, k \cdot p_2 \right. \right. \nonumber \\
&+& E_1\, E_2\, \left(3 m_e^2 + 2\, k \cdot p_1 - 4\, |{\bvec{p}_1}|\,
|{\bvec{p}_2}|\, x_1\, x_2 \right) \nonumber \\
&+& k \cdot p_1\, \left(4\, m_e^2 + 4\, k \cdot p_2 - 2\, |{\bvec{p}_1}|\,
|{\bvec{p}_2}|\, x_1\, x_2 \right) \nonumber \\
&+& \left. |{\bvec{p}_1}|\, |{\bvec{p}_2}|\, \left(-3\, m_e^2\, x_1\, x_2 +
|{\bvec{p}_1}|\, |{\bvec{p}_2}|\, \left(1 - x_1^2 + \left(-1 + 3\, x_1^2
\right)\, x_2^2 \right) \right) \right] \nonumber \\
&+& {{\ca}}^{\!\!\! \prime  2}\, \left[ -m_e^4 + 2\, E_1^2\, E_2^2 + 2\, (k
\cdot p_1)^2 - 3\, m_e^2\, k \cdot p_2 \right. \nonumber \\
&+& E_1\, E_2\, \left(-m_e^2 + 2\, k \cdot p_1 - 4\, |{\bvec{p}_1}|\,
|{\bvec{p}_2}|\, x_1\, x_2 \right) \nonumber \\
&+& k \cdot p_1\, \left(-m_e^2 + 4\, k \cdot p_2 - 2\, |{\bvec{p}_1}|\,
|{\bvec{p}_2}|\, x_1\, x_2 \right) \nonumber \\
&+& \left. |{\bvec{p}_1}|\, |{\bvec{p}_2}|\, \left( m_e^2\, x_1\, x_2 +
|{\bvec{p}_1}|\, |{\bvec{p}_2}|\, \left(1 - x_1^2 + \left(-1 + 3\, x_1^2
\right)\, x_2^2 \right) \right) \right] \nonumber \\
&-& 2~ \frac{\left(k \cdot p_1 - m_e^2 \right)\, k \cdot (p_1 + p_2)}
{{\sqrt{\frac{pole-1} {pole+1}}}} \nonumber \\
&\times& \left[-m_e^2 - E_1\, E_2 + k \cdot (p_1 + p_2) + |{\bvec{p}_1}|\,
|{\bvec{p}_2}|\, x_1\, x_2 - |{\bvec{p}_1}|\, |{\bvec{p}_2}|\, {\sqrt{1 -
{x_1}^2}}\, {\sqrt{1 - {x_2}^2}} \right]^{-1} \nonumber \\
&\times& \left. \left[{{\ca}}^{\!\!\! \prime  2}\, \left(-m_e^2 + k \cdot (p_1 +
2\, p_2) \right) + {{\cv}}^{\!\!\! \prime  2}\, \left(2 m_e^2 + k \cdot
(p_1 + 2\, p_2) \right) \right] \right\}
\eea where
\be
pole = \frac{m_e^2+E_1\, E_2\, - k\cdot (p_1+p_2) - |{\bvec{p}_1}|\,
|{\bvec{p}_2}|\, x_1\, x_2} {|{\bvec{p}_1}|\, |{\bvec{p}_2}|\,
\sqrt{(1-x_1^2)\, (1-x_2^2)}}\,\,\, ,
\ee
\item $d_2 \neq 0$, $\left| \frac{d_1}{d_2} \right| < 1$;
\bea
I_{F_1}
&=& \frac{256\, \pi}{m_e^2-k\cdot p_1}\, \left\{ {{\cv}}^{\!\!\! \prime  2}
\left[ 2\, E_1^2\, E_2^2 + 2\, (k \cdot p_1)^2 - 4\, m_e^2\, k \cdot p_2
-3\, m_e^4 \right. \right. \nonumber \\
&+& E_1\, E_2\, \left(2\, k \cdot p_1 + 3\, m_e^2 - 4\, |{\bvec{p}_1}|\,
|{\bvec{p}_2}|\, x_1\, x_2 \right) \nonumber \\
&+& k \cdot p_1\, \left(4\, k \cdot p_2 + 4\, m_e^2 - 2\, |{\bvec{p}_1}|\,
|{\bvec{p}_2}|\, x_1\, x_2 \right) \nonumber \\
&+& \left. |{\bvec{p}_1}|\, |{\bvec{p}_2}|\, \left(-3\, m_e^2\, x_1\, x_2 +
|{\bvec{p}_1}|\, |{\bvec{p}_2}|\, \left(1 - x_1^2 + \left(-1 + 3\, x_1^2
\right)\, x_2^2 \right) \right) \right] \nonumber \\
&+& {{\ca}}^{\!\!\! \prime  2} \left[ 2\, E_1^2\, E_2^2 + 2\, (k \cdot p_1)^2
- 3\, k \cdot p_2\, m_e^2 - m_e^4 \right. \nonumber \\
&+& E_1\, E_2\, \left(2\, k \cdot p_1 - m_e^2 - 4\, |{\bvec{p}_1}|\,
|{\bvec{p}_2}|\, x_1\, x_2 \right) \nonumber \\
&+& k \cdot p_1\, \left(4\, k \cdot p_2 - m_e^2 - 2\, |{\bvec{p}_1}|\,
|{\bvec{p}_2}|\, x_1\, x_2 \right) \nonumber \\
&+& \left. \left. |{\bvec{p}_1}|\, |{\bvec{p}_2}|\, \left( m_e^2\, x_1\, x_2 +
|{\bvec{p}_1}|\, |{\bvec{p}_2}|\, \left(1 - x_1^2 + \left(-1 + 3\, x_1^2
\right)\, x_2^2 \right) \right) \right] \right\}
\eea
\end{enumerate}

The functions $I_{F_2}$ in the same three cases above are obtained from
the previous ones with the substitutions $k\rightarrow -k$ and $p_1
\leftrightarrow p_2$.

\section{Bremsstrahlung functions}
\label{bremapp}
\setcounter{equation}0

The functions entering in Eq.(\ref{tem}) are the following:
\bea
\Phi^E_{1} &=& 512\, \left\{ {{\cv}}^{\!\!\! \prime  2}\, \left[ \left(k
\cdot p_1\right)^2\, \left(p_1 \cdot p_2 -2\, \left(m_e^2 +
k \cdot p_2 \right) \right) \right. \right. \nonumber \\
&-& 2\, m_e^2\, \left(m_e^2 - k \cdot p_2 + p_1 \cdot p_2 \right)\,
\left(2\, m_e^2 - k \cdot p_2 + p_1 \cdot p_2 \right) \nonumber \\
&+& \left. 2\, k \cdot p_1\, \left(4\, m_e^4 - (k \cdot p_2)^2 + 3\, m_e^2\,
p_1 \cdot p_2 + k \cdot p_2\, \left(p_1 \cdot p_2 -2\, m_e^2 \right)
\right) \right] \nonumber \\
&+& {{\ca}}^{\!\!\! \prime  2}\, \left[ 2\, m_e^6 - 2\, m_e^2\,
{\left((k-p_2) \cdot p_1\right)}^2 + (k \cdot p_1)^2\, \left( 4\, m_e^2
- 2\, k \cdot p_2 + p_1 \cdot p_2 \right) \right. \nonumber \\
&-& \left. \left. 2\, k \cdot p_1\, \left(2\, m_e^4 + (k \cdot p_2)^2
- k \cdot p_2\, \left(m_e^2 + p_1 \cdot p_2 \right) \right) \right]
\right\}\,\,\, , \\
\Phi^E_{2} &=& 512\, \left\{ {{\cv}}^{\!\!\! \prime 2}\, \left[ -2\,
\left(2\, m_e^6 - 4\, m_e^4\, k \cdot p_2 + (k \cdot p_1)^2\, \left(m_e^2
+ k \cdot p_2 \right) + m_e^2\, (k \cdot p_2)^2 \right)
\right. \right. \nonumber \\
&+& 2\, k \cdot p_1\, \left(3\, m_e^4 - \left(2\, m_e^2 + k \cdot p_2 \right)\,
\left(k \cdotp_2 - p_1 \cdot p_2 \right) \right) \nonumber \\
&-& \left. \left(6\, m_e^4 - 6\, m_e^2\, k \cdot p_2 - (k \cdot p_2)^2 \right)\,
p_1 \cdot p_2 - 2\, m_e^2\, \left(  p_1 \cdot p_2 \right)^2 \right] \nonumber \\
&+& {{\ca}}^{\!\!\! \prime  2}\, \left[ 2\, m_e^6 - 2\, (k \cdot p_1)^2\,
\left(m_e^2 + k \cdot p_2 \right) - 2\, k \cdot p_2\, \left(2\, m_e^4 +
k \cdot p_1\, \left(k \cdot p_2 - m_e^2\right) \right) \right. \nonumber \\
&+& 4\, m_e^2\, (k \cdot p_2)^2 + \left(2\, k \cdot p_1\, \left(2\, m_e^2 +
k \cdot p_2 \right) + (k \cdot p_2)^2 \right)\, p_1 \cdot p_2 \nonumber \\
&-& \left. \left. 2\, m_e^2\, \left(  p_1 \cdot p_2 \right)^2 \right]
\right\}\,\,\, , \\
\Phi^E_{12} &=& 512\, \left\{ {{\cv}}^{\!\!\! \prime  2}\, \left[ 2\,
m_e^4\, k \cdot p_2 - m_e^2\, \left(k \cdot p_2 \right)^2 - (k \cdot
p_1)^2\, \left(m_e^2 - 2\, p_1 \cdot p_2 \right) \right. \right. \nonumber \\
&-& 2\, m_e^4\, p_1 \cdot p_2 + 2\, m_e^2\, k \cdot p_2\, p_1 \cdot p_2 +
2\, \left(k \cdot p_2 \right)^2\, p_1 \cdot p_2 \nonumber \\
&+& k \cdot p_1\, \left(k \cdot p_2\, \left(3\, p_1 \cdot p_2 - 4\, m_e^2
\right) + 2\, \left(m_e^4 + m_e^2\, p_1 \cdot p_2 - 2\, (p_1 \cdot
p_2)^2 \right) \right) \nonumber \\
&-& \left. 4\, k \cdot p_2\, (p_1 \cdot p_2)^2 + 2\, (p_1 \cdot p_2)^3
\right] \nonumber \\
&+& {{\ca}}^{\!\!\! \prime  2}\, \left[ - m_e^4\, k \cdot p_2 + m_e^2\,
\left(k \cdot p_2 \right)^2 + 4\, m_e^4\, p_1 \cdot p_2 - 7\, m_e^2\,
k \cdot p_2\, p_1 \cdot p_2 \right. \nonumber \\
&+& 2\, \left(k \cdot p_2 \right)^2\, p_1 \cdot p_2 + (k \cdot p_1)^2\,
\left(m_e^2 + 2\, p_1 \cdot p_2 \right) + 2\, \left(3\, m_e^2 - 2\,
k \cdot p_2 \right)\, (p_1 \cdot p_2)^2 \nonumber \\
&-& \left. \left. k \cdot p_1\, \left(m_e^4 + \left(7\, m_e^2 - 3\, k \cdot
p_2 \right)\, p_1 \cdot p_2 + 4\, (p_1 \cdot p_2)^2 \right) + 2\, (p_1
\cdot p_2)^3 \right] \right\}\,\,\, .
\eea
The corresponding functions appearing in Eq.(\ref{tam}) are, instead,
obtained from the above reported ones by replacing the 4-momentum $k$ with
$-k$.

After integration over the angle $\phi$ as described in Section
\ref{bremcor}, the functions $I_a$ for the photon absorption and $I_{e0},
I_e$ for the photon emission, appearing in Eq.(\ref{qeqa}), are the
following($E_{\nu\bar{\nu}}^\pm = E_1+E_2\pm |\bvec{k}|$, $x_{1,2} = \cos
\theta_{1,2}$, $s_{1,2} = \sin \theta_{1,2}$, $\epsilon^2=E_1\, E_2 -
|{\bvec{p}_1}|\, |{\bvec{p}_2}|\, x_1\, x_2$, $\mu_{1,2} = E_{1,2} -
|{\bvec{p}_{1,2}}| x_{1,2}$)
\bea
I_a
&=& \frac{128\, \pi\, |\bvec{k}|}{\mu_1^2\, \mu_2^2} \left\{ 3\, \ca^{\!\!\!
  \prime  2}\, E_{\nu\bar{\nu}}^+\, \epsilon^2\, \mu_1\, \mu_2^3 \right.
  \nonumber \\
&-& \cv^{\!\!\! \prime  2}\, \left[ 3\, m_e^4\, \mu_2^3 + m_e^2\, \mu_1^3\,
  (3\, m_e^2 + E_{\nu\bar{\nu}}^+\, \mu_2) + m_e^2\, \mu_1^2\, \mu_2\, (3\,
  m_e^2 - 6\, \epsilon^2 + 2\, E_{\nu\bar{\nu}}^+\, \mu_2) \right. \nonumber
  \\
&+& \left. \mu_1\, \mu_2^2\, \left(3\, m_e^4 + m_e^2\, E_{\nu\bar{\nu}}^+\,
  \mu_2 - 3\, \epsilon^2\, (2\, m_e^2 + E_{\nu\bar{\nu}}^+\, \mu_2)
  \right) \right] \nonumber \\
&+& \left({{\ca}}^{\!\!\! \prime  2} + {{\cv}}^{\!\!\! \prime  2} \right)\,
\left[ 4\,
  \epsilon^4\, \mu_1\, \mu_2\, \left( \mu_1 + \mu_2 \right) \right. \nonumber
  \\
&+& E_{\nu\bar{\nu}}^+\, \left(- m_e^2\, \mu_1^4 + |{\bvec{k}}|\, \mu_1\,
  \mu_2\, (\mu_1^3 + \mu_1^2\, \mu_2 + \mu_1\, \mu_2^2) + \mu_2^4\,
  (|{\bvec{k}|}\, \mu_1 - m_e^2 ) \right) \nonumber \\
&+& \epsilon^2\, \left(-2\, m_e^2\, \mu_1\, \mu_2^2 - 2\, m_e^2\, \mu_2^3 +
  2\, \mu_1^2\, \mu_2\, (2\, E_{\nu\bar{\nu}}^+\, \mu_2 - m_e^2) + \mu_1^3\,
  (3\, E_{\nu\bar{\nu}}^+\, \mu_2 - 2\, m_e^2) \right) \nonumber \\
&+& \left. \left. 2\, | {\bvec{p}_1} | ^2\, | {\bvec{p}_2} | ^2\, \mu_1\,
\mu_2\, (\mu_1 + \mu_2)\, (1 - x_1^2)\, (1 - x_2^2) \right] \right\}\,\,\,
,
\label{fa}
\eea
and
\bea
I_{e(e0)} &=& \frac{-\, 64\, \pi\, |\bvec{k}|}{\mu_1^2\, \mu_2^2}\, (2\,
\eta_1\, + \eta_3)\,\,\, , \quad\quad\quad\mbox{($a_1 =0$, $a_2\geq 0$ or
$a_1 \neq 0$, $a_2/a_1 \geq 1$)}
\label{fe1} \\
I_{e(e0)} &=& \frac{-\, 64\, |\bvec{k}|}{\mu_1^2\, \mu_2^2}\, \left[ 2\,
(\pi  - \alpha)\, \eta_1 - 2\, \sin\alpha\, \eta_2 + (\pi  - \alpha  -
\cos\alpha\, \sin\alpha)\, \eta_3 \right. \nonumber \\
&-& \left. \frac{9\, \sin\alpha + \sin 3\, \alpha}{6}~ \eta_4
\right]\,\,\, , \quad\quad\quad\mbox{($a_1 \neq 0$, $-1 < a_2/a_1
< 1$)}
\label{fe2}
\eea
where $a_1$, $a_2$, and $\alpha$ have been defined in Section
\ref{bremcor} and we have introduced the following functions ($\zeta=0$
for $I_e$ while $\zeta=1$ for $I_{e0}$):
\bea \eta_1 &=& \cv^{\!\!\! \prime  2}\, \left[ 3\,
m_e^4\, (\mu_1 + \mu_2)\, (\mu_1^2
  + \mu_2^2) \right. \nonumber \\
&+& (\mu_1 + \mu_2)\, \left(-4\, E_1^2\, E_2^2\, \mu_1\, \mu_2 +
  E_{\nu\bar{\nu}}^-\, |\bvec{k}|\, \mu_1\, \mu_2\, (\mu_1^2 + \mu_2^2) \right)
  \nonumber \\
&+& \left. m_e^2\, \left(2\, \epsilon^2\, (\mu_1 + \mu_2)\, (\mu_1^2 - 3\,
  \mu_1\, \mu_2 + \mu_2^2) + E_{\nu\bar{\nu}}^-\, (\mu_1^2 + \mu_2^2)\,
  (\mu_1^2 + \mu_1\, \mu_2 + \mu_2^2) \right) \right] \nonumber \\
&+& \ca^{\!\!\! \prime  2}\, \left[ \mu_1\, \mu_2\, (\mu_1 + \mu_2)\,
  \left(-4\, E_1^2\, E_2^2 + E_{\nu\bar{\nu}}^-\, |\bvec{k}|\, (\mu_1^2 +
  \mu_2^2) \right) \right. \nonumber \\
&+& \left. m_e^2\, \left(2\, \epsilon^2\, (\mu_1 + \mu_2)\, (\mu_1^2 + \mu_2^2)
  + E_{\nu\bar{\nu}}^-\, (\mu_1^4 + \mu_2^4) \right) \right] \nonumber \\
&+& \left({{\ca}}^{\!\!\! \prime  2} + {{\cv}}^{\!\!\! \prime  2} \right)\,
  \mu_1\, \mu_2\, \left[ - E_{\nu\bar{\nu}}^-\, \epsilon^2\, \left(3\, \mu_1^2
  + 4\, \mu_1\, \mu_2 + 3\, \mu_2^2 \right) \right. \nonumber \\
&+& \left. 4\, \left( E_1\, E_2 + \epsilon^2 \right)\, \left( \mu_1 + \mu_2
  \right)\, |{\bvec{p}_1}| \, |{\bvec{p}_2}| \, x_1\, x_2 \right] \nonumber \\
&+& \zeta\, \frac{2\, \epsilon^2\, \mu_1\, \mu_2 - m_e^2\, (\mu_1^2 + \mu_2^2)}
  {|{\bvec{k}}|^2}\, \left\{ \cv^{\!\!\! \prime  2}\, \left[- E_{\nu\bar{\nu}}^-\,
  \left(\epsilon^4 + 3\, m_e^2\, \epsilon^2 + 2\, m_e^4 \right) \right. \right.
  \nonumber \\
&+& \left. 3\, m_e^2\, |{\bvec{k}}|\, (E_1 + E_2)\, (\mu_1 + \mu_2) \right]
  \nonumber \\
&+& \ca^{\!\!\! \prime  2}\, \left[ - E_{\nu\bar{\nu}}^-\, \left( E_1^2\, E_2^2 -
  m_e^4 \right) + E_{\nu\bar{\nu}}^-\, \left( E_1\, E_2 + \epsilon^2 \right)\,
  |{\bvec{p}_1}|\, |{\bvec{p}_2}|\, x_1\, x_2 \right] \nonumber \\
&+& \left. 2\, \left(\ca^{\!\!\! \prime  2} + \cv^{\!\!\! \prime  2} \right)\,
  \epsilon^2\, |{\bvec{k}}|\, (E_1 + E_2)\, \left( \mu_1 + \mu_2 \right)
  \right\}\,\,\, , \\
\eta_2 &=& |{\bvec{p}_1}|\, |{\bvec{p}_2}|\, s_1\, s_2\, \left\{6\, \cv^{\!\!\!
  \prime  2}\, m_e^2\, \mu_1\, \mu_2\, (\mu_1 + \mu_2) \right. \nonumber \\
&+& \left(\ca^{\!\!\! \prime  2} + \cv^{\!\!\! \prime  2} \right)\, \left[ -2\,
  m_e^2\, (\mu_1 + \mu_2)\, (\mu_1^2 + \mu_2^2) + \mu_1\, \mu_2\, \left( 8\,
  \epsilon^2\, (\mu_1 + \mu_2) \right. \right. \nonumber \\
&+& \left. \left. \left. E_{\nu\bar{\nu}}^-\, (3\, \mu_1^2 + 4\, \mu_1\,
  \mu_2 + 3\, \mu_2^2) \right) \right] \right\} \nonumber \\
&+& \zeta~ \frac{|{\bvec{p}_1}|\, |{\bvec{p}_2}|\, s_1\, s_2}{|{\bvec{k}}|^2}\,
  \left\{ -2\, \ca^{\!\!\! \prime  2}\, m_e^4\, E_{\nu\bar{\nu}}^-\, \mu_1\,
  \mu_2 \right. \nonumber \\
&+& \cv^{\!\!\! \prime  2}\, \left[ -6\, m_e^2\,(E_1 + E_2)\,
  |{\bvec{k}}|\, \mu_1\, \mu_2\, (\mu_1 + \mu_2) + E_{\nu\bar{\nu}}^-\,
  \left(12\, m_e^2\, \epsilon^2\, \mu_1\, \mu_2 \right. \right. \nonumber \\
&+& \left. \left. m_e^4\, (-3\, \mu_1^2 + 4\, \mu_1\, \mu_2 - 3\,
  \mu_2^2) \right) \right] \nonumber \\
&+& \left(\ca^{\!\!\! \prime  2} + \cv^{\!\!\! \prime  2} \right)\,
  \left[ 2\, E_{\nu\bar{\nu}}^-\, \epsilon^2\, \left( 3\, \epsilon^2\, \mu_1\,
  \mu_2 - m_e^2\, (\mu_1^2 + \mu_2^2) \right) \right. \nonumber \\
&+& \left. \left. 2\, (E_1 + E_2)\, |{\bvec{k}}|\, (\mu_1 + \mu_2)\, \left(
  -4\, \epsilon^2\, \mu_1\, \mu_2 + m_e^2\, (\mu_1^2 + \mu_2^2) \right) \right]
  \right\}\,\,\, , \\
\eta_3
&=& -\, 4\, \left(\ca^{\!\!\! \prime  2} + \cv^{\!\!\! \prime  2} \right)\,
  |{\bvec{p}_1}|^2\, |{\bvec{p}_2}|^2\, \mu_1\, \mu_2\, (\mu_1 + \mu_2)\, s_1^2\,
  s_2^2 \nonumber \\
&+& \zeta\, \frac{|{\bvec{p}_1}|^2\, |{\bvec{p}_2}|^2\, s_1^2\, s_2^2}
  {|{\bvec{k}}|^2}\, \left\{ -\, 6\, \cv^{\!\!\! \prime  2}\, m_e^2\,
  E_{\nu\bar{\nu}}^-\, \mu_1\, \mu_2 \right. \nonumber \\
&+& \left(\ca^{\!\!\! \prime  2} + \cv^{\!\!\! \prime  2} \right)\, \left[ -\,
  6\, E_{\nu\bar{\nu}}^-\, \epsilon^2\, \mu_1\, \mu_2 + 4\, (E_1 + E_2)\,
  |{\bvec{k}}|\, \mu_1\, \mu_2\, (\mu_1 + \mu_2) \right. \nonumber \\
&+& \left. \left. m_e^2\, E_{\nu\bar{\nu}}^-\, (\mu_1^2 + \mu_2^2) \right]
  \right\}\,\,\, , \\
\eta_4 &=& \zeta~ \frac{2\, \left(\ca^{\!\!\! \prime  2} + \cv^{\!\!\! \prime  2}
  \right)\, E_{\nu\bar{\nu}}^-\, |{\bvec{p}_1}|^3\, |{\bvec{p}_2}|^3\,
  \mu_1\, \mu_2\, s_1^3\, s_2^3}{|{\bvec{k}}|^2}\,\,\, .
\eea

\section{Photoproduction functions}
\label{photoapp}

The functions entering in Eq.(\ref{tmph}) are given by
\bea
\Phi_{\mathrm ph}^{1}
&=& \cv^{\!\!\! \prime  2}\, \left\{-32\, (k \cdot p_1)^3\, m_e^2 +
  4\, k \cdot p_1\, (8\, m_e^4\, m_\gamma^2 + 3\, m_e^2\, m_\gamma^4)
  \right. \nonumber \\
&+& 4\, (k \cdot p_1)^2\, \left[ -8\, m_e^4 + k \cdot p_2\, (8\, m_e^2
  + 6\, m_\gamma^2) + 7\, m_\gamma^2\, p_1 \cdot p_2 \right. \nonumber \\
&+& \left. m_e^2\, (-7\, m_\gamma^2 + 12\, p_1 \cdot p_2) \right] +
  m_\gamma^2\, \left[ 32\, m_e^6 - 16\, k \cdot p_2\, (3\, m_e^4 + 2\,
  m_e^2\, m_\gamma^2) \right. \nonumber \\
&+& \left. \left. 8\, m_e^4\, (5\, m_\gamma^2 - 6\, p_1 \cdot p_2) -
  m_\gamma^4\, p_1 \cdot p_2 + 9\, m_e^2\, (m_\gamma^4 - 4\, m_\gamma^2\,
  p_1 \cdot p_2) \right] \right\} \nonumber \\
&+& \ca^{\!\!\! \prime  2}\, \left\{16\, (k \cdot p_1)^3\, m_e^2 - 8\,
  k \cdot p_1\, (2\, m_e^4\, m_\gamma^2 + 3\, m_e^2\, m_\gamma^4) \right.
  \nonumber \\
&-& 4\, (k \cdot p_1)^2\, \left(4\, k \cdot p_2\, m_e^2 - 4\, m_e^4 - 6\,
  k \cdot p_2\, m_\gamma^2 + m_e^2\, m_\gamma^2 - 7\, m_\gamma^2\, p_1
  \cdot p_2 \right) \nonumber \\
&-& \left. m_\gamma^2\, \left[ m_e^2\, \left(16\, m_e^4 - 4\, (k \cdot p_2
  - 5\, m_e^2)\, m_\gamma^2 + 9\, m_\gamma^4 \right) + m_\gamma^4\, p_1
  \cdot p_2 \right] \right\} \nonumber \\
&+& 2\, \left(\ca^{\!\!\! \prime  2} + \cv^{\!\!\! \prime  2} \right)\,
  \left\{-8\, (k \cdot p_1)^2\, p_1 \cdot p_2\, (k \cdot p_2 + p_1 \cdot
  p_2) \right. \nonumber \\
&+& 4\, (k \cdot p_1)^3\, (k \cdot p_2 + 2\, p_1 \cdot p_2) + 2\,
  m_\gamma^2\, (k \cdot p_2 + p_1 \cdot p_2)\, \left(4\, k \cdot p_2\,
  m_e^2 \right. \nonumber \\
&+& \left. (4\, m_e^2 + m_\gamma^2)\, p_1 \cdot p_2 \right) + k \cdot p_1\,
  m_\gamma^2\, \left(-8\, (k \cdot p_2)^2 \right. \nonumber \\
&+& \left. \left. k \cdot p_2\, (5\, m_\gamma^2 - 16\, p_1 \cdot p_2) + 4\,
  (m_\gamma^2 - 2\, p_1 \cdot p_2)\, p_1 \cdot p_2 \right) \right\}\,\,\, ,
\\
\Phi_{\mathrm ph}^{2}
&=& \cv^{\!\!\! \prime  2}\, \left\{ -8\, k \cdot p_1\, \left[4\, (k \cdot
  p_2)^2\, m_e^2 + 3\, \left((k \cdot p_2)^2 - 2\, m_e^4 \right)\, m_\gamma^2
  - 4\, m_e^2\, m_\gamma^4 \right] \right. \nonumber \\
&+& m_e^2\, \left[ 32\, (k \cdot p_2)^2\, (k \cdot p_2 - m_e^2) - 4\, \left(7\,
  (k \cdot p_2)^2 + 8\, k \cdot p_2\, m_e^2 - 8\, m_e^4 \right)\, m_\gamma^2
  \right. \nonumber \\
&-& \left. 4\, (3\, k \cdot p_2 - 10\, m_e^2)\, m_\gamma^4 + 9\, m_\gamma^6
  \right] - \left[ 48\, m_e^4\, m_\gamma^2 + 36\, m_e^2\, m_\gamma^4 +
  m_\gamma^6 \right. \nonumber \\
&-& \left. \left. 4\, (k \cdot p_2)^2\, (12\, m_e^2 + 7\, m_\gamma^2) \right]\,
  p_1 \cdot p_2 \right\} \nonumber \\
&+& \ca^{\!\!\! \prime  2}\, \left\{ m_e^2\, \left[ -16\, (k \cdot p_2)^2\,
  (k \cdot p_2 - m_e^2) - 4\, (k \cdot p_2 - 2\, m_e^2)^2\, m_\gamma^2 \right.
  \right. \nonumber \\
&+& \left. 4\, (6\, k \cdot p_2 - 5\, m_e^2)\, m_\gamma^4 - 9\, m_\gamma^6
  \right] - 4\, k \cdot p_1\, \left(m_e^2\, m_\gamma^4 \right. \nonumber \\
&+& \left. \left. (k \cdot p_2)^2\, (-4\, m_e^2 + 6\, m_\gamma^2) \right) -
  m_\gamma^2\, \left(-28\, (k \cdot p_2)^2 + m_\gamma^4 \right)\, p_1 \cdot
  p_2 \right\} \nonumber \\
&+& 2\, \left(\ca^{\!\!\! \prime  2} + \cv^{\!\!\! \prime  2} \right)\, \left\{
  k \cdot p_1\, \left[ 4\, (k \cdot p_2)^3 + 8\, k \cdot p_1\, (k \cdot p_2 +
  m_e^2)\, m_\gamma^2 \right. \right. \nonumber \\
&+& \left. 5\, k \cdot p_2\, m_\gamma^4 \right] + 2\, \left[4\, k \cdot (p_1 -
  p_2)\, (k \cdot p_2)^2 - 8\, k \cdot p_1\, (k \cdot p_2 + m_e^2)\, m_\gamma^2
  \right. \nonumber \\
&-& \left. (k \cdot p_1 + 2\, k \cdot p_2)\, m_\gamma^4 \right]\, p_1 \cdot p_2
  - 2\, \left[4\, (k \cdot p_2)^2 \right. \nonumber \\
&-& \left. \left. 4\, (k \cdot p_2 + m_e^2)\, m_\gamma^2 - m_\gamma^4 \right]\,
  (p_1 \cdot p_2)^2 \right\}\,\,\, , \\
\Phi_{\mathrm ph}^{12}
&=&  \cv^{\!\!\! \prime  2}\, \left\{ -4\, (k \cdot p_1)^2\, \left(8\, k \cdot
  p_2\, m_e^2 + (k \cdot p_2 - 2\, m_e^2)\, m_\gamma^2 \right) \right. \nonumber
  \\
&+& m_\gamma^2\, \left[ 8\, (k \cdot p_2)^2\, m_e^2 + 32\, m_e^4\, p_1 \cdot p_2
  + 5\, m_\gamma^2\, (m_\gamma^2 - 4\, p_1 \cdot p_2)\, p_1 \cdot p_2 \right.
  \nonumber \\
&+& m_e^2\, \left(3\, m_\gamma^4 + 28\, m_\gamma^2\, p_1 \cdot p_2 - 48\, (p_1
  \cdot p_2)^2 \right) - 2\, k \cdot p_2\, \left( 4\, m_e^4 \right. \nonumber \\
&+& \left. \left. 7\, m_\gamma^2\, p_1 \cdot p_2 + m_e^2\, (5\, m_\gamma^2 +
  16\, p_1 \cdot p_2) \right) \right] + 2\, k \cdot p_1\, \left[ 2\, (k \cdot
  p_2)^2\, (8\, m_e^2 + m_\gamma^2) \right. \nonumber \\
&-& k \cdot p_2\, \left( 16\, m_e^4 + m_\gamma^4 + m_e^2\, (22\, m_\gamma^2 -
  24\, p_1 \cdot p_2) + 2\, m_\gamma^2\, p_1 \cdot p_2 \right) \nonumber \\
&+& \left. \left. m_\gamma^2\, \left( 4\, m_e^4 + 7\, m_\gamma^2\, p_1 \cdot p_2
  + m_e^2\, (5\, m_\gamma^2 + 16\, p_1 \cdot p_2) \right) \right] \right\}
  \nonumber \\
&+& \ca^{\!\!\! \prime  2}\, \left\{ 4\, (k \cdot p_1)^2\, k \cdot p_2\, (4\,
  m_e^2 - m_\gamma^2) + m_\gamma^2\, \left[ -3\, m_e^2\, m_\gamma^4 \right.
  \right. \nonumber \\
&+& 4\, m_e^4\, (m_\gamma^2 - 4\, p_1 \cdot p_2) + 5\, m_\gamma^2\, (m_\gamma^2 -
  4\, p_1 \cdot p_2)\, p_1 \cdot p_2 \nonumber \\
&-& \left. 2\, k \cdot p_2\, \left( 4\, m_e^4 + 7\, m_\gamma^2\, p_1 \cdot p_2 -
  m_e^2\, (m_\gamma^2 + 8\, p_1 \cdot p_2) \right) \right] \nonumber \\
&-& 2\, k \cdot p_1\, \left[ (k \cdot p_2)^2\, (8\, m_e^2 - 2\, m_\gamma^2) +
  k \cdot p_2\, (-8\, m_e^4 + 2\, m_e^2\, m_\gamma^2 + m_\gamma^4 \right.
  \nonumber \\
&+& \left. \left. 2\, m_\gamma^2\, p_1 \cdot p_2) + m_\gamma^2\, \left(-4\, m_e^4
  - 7\, m_\gamma^2\, p_1 \cdot p_2 + m_e^2\, (m_\gamma^2 + 8\, p_1 \cdot p_2)
  \right) \right] \right\} \nonumber \\
&+& 8\, \left(\ca^{\!\!\! \prime  2} + \cv^{\!\!\! \prime  2} \right)\, \left\{
  m_\gamma^2\, p_1 \cdot p_2\, (k \cdot p_2 + p_1 \cdot p_2)\, (k \cdot p_2 +
  2\, p_1 \cdot p_2) \right. \nonumber \\
&+& (k \cdot p_1)^2\, \left((k \cdot p_2)^2 + 2\, k \cdot p_2\, p_1 \cdot p_2 +
  m_\gamma^2\, p_1 \cdot p_2 \right) \nonumber \\
&-& \left. k \cdot p_1\, p_1 \cdot p_2\, \left(2\, (k \cdot p_2)^2 + 2\, k \cdot
  p_2\, p_1 \cdot p_2 + 3\, m_\gamma^2\, p_1 \cdot p_2 \right) \right\}\,\,\, .
\eea
The function $I_{\mathrm ph}$ appearing in Eq.(\ref{qphom}) is obtained
after integration over the angle $\phi$. In particular, one has
\bea
I_{\mathrm ph} &=& \frac{128\, \pi\, |{\bvec{k}}|^2\, (E_1 - E_2 +
E_\gamma)}{E_\gamma\, (m_\gamma^2+2\, k \cdot p_1)^2\, (m_\gamma^2-2\, k
\cdot p_1)^2}~ (2\, \nu_1 + \nu_3) \quad\quad b_1 = 0,~
b_2 \leq 0 \,\,\, , \\
I_{\mathrm ph} &=& \frac{128\, |{\bvec{k}}|^2\, (E_1 - E_2 +
E_\gamma)}{E_\gamma\, (m_\gamma^2+2\, k \cdot p_1)^2\, (m_\gamma^2-2\, k
\cdot p_1)^2}~ \left[2\, \beta\, \nu_1 + 2\, \sin\beta\, \nu_2 + \left(
\beta + \cos\beta\, \sin\beta \right)\, \nu_3 \right. \nonumber \\
&+& \left. \frac{9\, \sin\beta + \sin\, 3\beta}{6}~ \nu_4 \right]
\quad\quad\quad\quad\quad\quad\quad\quad\quad\quad b_1 \neq 0,~ -1 <
b_2/b_1 <1\,\,\, ,
\eea
and the functions $\nu_i$ are ($x_{1,2} = \cos \theta_{1,2}$, $s_{1,2} =
\sin \theta_{1,2}$, $\epsilon^2=E_1\, E_2 - |{\bvec{p}_1}|\,
|{\bvec{p}_2}|\, x_1\, x_2$)
\bea
\nu_1
&=& \ca^{\!\!\! \prime  2}\, \left\{ -8\, k \cdot p_1\, k \cdot p_2\,
  \left[ 2\, \epsilon^6 - 4\, \epsilon^4\, k \cdot (p_1 - p_2) - k \cdot
  (p_1 - p_2)\, \left( (k \cdot p_1)^2 + (k \cdot p_2)^2 \right) \right.
  \right. \nonumber \\
&+& \left. \epsilon^2\, \left( 3\, (k \cdot p_1)^2 - 4\, k \cdot p_1\,
  k \cdot p_2 + 3\, (k \cdot p_2)^2 \right) \right] \nonumber \\
&+& 8\, m_e^2\, \left[ (k \cdot p_1)^4 + \epsilon^2\, \left( \epsilon^2 - 2\, k
  \cdot (p_1 - p_2) \right)\, \left( (k \cdot p_1)^2 + (k \cdot p_2)^2
  \right) + (k \cdot p_2)^4 \right] \nonumber \\
&+& 16\, m_e^4\, \epsilon^2\, k \cdot p_1\, k \cdot p_2 - 8\, m_e^6\, \left( (k
  \cdot p_1)^2 + (k \cdot p_2)^2 \right) \nonumber \\
&+& 4\, m_\gamma^2\, \left[ -3\, \epsilon^4\, (k \cdot p_1)^2 + 2\, \epsilon^6\,
  k \cdot (p_1 - p_2) + \epsilon^2\, k \cdot (p_1 - p_2)\, \left( (k \cdot p_1)^2
  + (k \cdot p_2)^2 \right) \right. \nonumber \\
&+& 14\, \epsilon^4\, k \cdot p_1\, k \cdot p_2 - 12\, \epsilon^2\, k \cdot p_1\,
  k \cdot p_2\, k \cdot (p_1 - p_2)\nonumber \\
&+& 3\, k \cdot p_1\, \left( (k \cdot p_1)^2 + (k \cdot p_2)^2 \right)\, k \cdot
  p_2 - 3\, \epsilon^4\, (k \cdot p_2)^2 - 4\, (k \cdot p_1)^2\, (k \cdot p_2)^2
  \nonumber \\
&+& 2\, m_e^2\, \left((k \cdot p_1)^3 + \epsilon^2\, k \cdot (p_1 - p_2) \,
  (\epsilon^2 - k \cdot (p_1 - p_2))\right. \nonumber \\
&-& \left. 2\, \epsilon^2\, \left( (k \cdot p_1)^2 + (k \cdot p_2)^2 \right) + 2\,
  k \cdot p_1\, k \cdot (p_1 - p_2)\, k \cdot p_2 - (k \cdot p_2)^3 \right)
  \nonumber \\
&-& \left. m_e^4\, k \cdot (p_1 - p_2)\, \left(2\, \epsilon^2 + k \cdot (p_1 - p_2)
  \right) - 2\, m_e^6\, k \cdot (p_1 - p_2) \right] \nonumber \\
&+& 2\, m_\gamma^4\, \left[ 2\, \epsilon^6 + 5\, \epsilon^2\, (k \cdot p_1)^2 -
  8\, \epsilon^4\, k \cdot (p_1 - p_2) - 18\, \epsilon^2\, k \cdot p_1\, k \cdot
  p_2 \right. \nonumber \\
&+& 4\, k \cdot p_1\, k \cdot p_2\, k \cdot (p_1 - p_2) + 5\, \epsilon^2\, (k
  \cdot p_2)^2  - 2\, m_e^6 + m_e^2\, \left( 2\, \epsilon^2\, (\epsilon^2 - 2\, k
  \cdot (p_1 - p_2)) \right. \nonumber \\
&-& \left. \left. 3\, \left( (k \cdot p_1)^2 + (k \cdot p_2)^2 \right) + 14\, k
  \cdot p_1\, k \cdot p_2 \right) - 2\, m_e^4\, \left( \epsilon^2 + 2\, k \cdot
  (p_1 - p_2) \right) \right] \nonumber \\
&-& 2\, m_\gamma^6\, \left[ 2\, \epsilon^4 - 3\, \epsilon^2\, k \cdot (p_1 - p_2)
  - k \cdot p_1\, k \cdot p_2 + 5\, m_e^2\, k \cdot (p_1 - p_2) + 2\, m_e^4 \right]
  \nonumber \\
&+& \left. m_\gamma^8\, (\epsilon^2 - 3\, m_e^2) \right\} \nonumber \\
&+& \cv^{\!\!\! \prime  2}\, \left[ 2\, m_e^2 + m_\gamma^2 - 2\, \epsilon^2 + 2\, k
  \cdot (p_1 - p_2) \right] \nonumber \\
&\times& \left\{2\, \epsilon^4\, \left(2\, k \cdot p_2 - m_\gamma^2 \right)\, \left(
  2\, k \cdot p_1 + m_\gamma^2 \right) + \epsilon^2\, \left[ -8\, k \cdot p_1\, k
  \cdot p_2\, k \cdot (p_1 - p_2) \right. \right. \nonumber \\
&-& 4\, m_e^2\, \left( \left( (k \cdot p_1)^2 + (k \cdot p_2)^2 \right) + 4\, k
  \cdot p_1\, k \cdot p_2 \right) + 2\, m_\gamma^2\, \left((k \cdot (p_1 - p_2))^2
  \right. \nonumber \\
&-& \left. \left. 6\, k \cdot p_1\, k \cdot p_2 + 2\, m_e^2\, k \cdot (p_1 - p_2)
  \right) + 2\, m_\gamma^4\, \left(2\, k \cdot (p_1 - p_2) + m_e^2 \right) +
  m_\gamma^6 \right] \nonumber \\
&+& \left[ 2\, \left( (k \cdot p_1)^2 + (k \cdot p_2)^2 \right) + 2\, m_\gamma^2\,
  k \cdot (p_1 - p_2) + m_\gamma^4 \right]\, \left[ 2\, k \cdot p_1\, \left( k \cdot
  p_2 + m_e^2 \right) \right. \nonumber \\
&+& \left. \left. m_e^2\, \left(-2\, k \cdot p_2 + 4\, m_e^2 + 3\, m_\gamma^2 \right)
  \right] \right\}\,\,\, ,\\
\nu_2
&=& |{\bvec{p}_1}|\, |{\bvec{p}_2}|\, s_1\, s_2 \left\{ \ca^{\!\!\! \prime  2}\,
  \left\{ 8\, k \cdot p_1\, k \cdot p_2\, \left[ - 8\, \epsilon^2\, k \cdot (p_1 -
  p_2) + 3\, \left( (k \cdot p_1)^2 + (k \cdot p_2)^2 \right) \right. \right. \right.
  \nonumber \\
&-& \left. 4\, k \cdot p_1\, k \cdot p_2 + 6\, \epsilon^4 \right] - 16\, m_e^2\,
  \left[ -(k \cdot p_1)^3 + \epsilon^2\, \left( (k \cdot p_1)^2 + (k \cdot p_2)^2
  \right) + (k \cdot p_2)^3 \right. \nonumber \\
&+& \left. k \cdot p_1\, k \cdot p_2\, k \cdot (p_1 - p_2) \right] - 16\, m_e^4\,
  k \cdot p_1\, k \cdot p_2 + m_\gamma^2\, \left[ -4\, \left( 6\, \epsilon^4\, k
  \cdot (p_1 - p_2) \right. \right. \nonumber \\
&-& 6\, \epsilon^2\, \left( (k \cdot p_1)^2 + (k \cdot p_2)^2 \right) + k \cdot p_1\,
  k \cdot p_2\, \left( 28\, \epsilon^2 - 13\, k \cdot (p_1 - p_2) \right) - (k \cdot
  p_2)^3 \nonumber \\
&+& \left. (k \cdot p_1)^3 \right) + 8\, m_e^2\, \left( -2\, \epsilon^2\, k \cdot
  (p_1 - p_2) + 3\, \left( (k \cdot p_1)^2 + (k \cdot p_2)^2 \right) - 2\, k \cdot
  p_1\, k \cdot p_2 \right) \nonumber \\
&+& \left. 8\, m_e^4\, k \cdot (p_1 - p_2) \right] - 2\, m_\gamma^4\, \left[ 6\,
  \epsilon^4 - 16\, \epsilon^2\, k \cdot (p_1 - p_2) + 5\, \left( (k \cdot p_1)^2
  + (k \cdot p_2)^2 \right) \right. \nonumber \\
&-& \left. 18\, k \cdot p_1\, k \cdot p_2 + 4\, m_e^2\, \left( \epsilon^2 - k \cdot
  (p_1 - p_2) \right) + 4\, m_e^4 \right] \nonumber \\
&+& \left. 2\, m_\gamma^6\, \left(4\, \epsilon^2 - 3\, k \cdot (p_1 - p_2) \right)
  - m_\gamma^8 \right\} \nonumber \\
&+& \cv^{\!\!\! \prime  2}\, \left\{ 8\, k \cdot p_1\, k \cdot p_2\, \left[ 6\,
  \epsilon^4 - 8\, \epsilon^2\, k \cdot (p_1 - p_2) + 3\, \left( (k \cdot p_1)^2 +
  (k \cdot p_2)^2 \right) \right. \right. \nonumber \\
&-& \left. 4\, k \cdot p_1\, k \cdot p_2 \right] - 16\, m_e^2\, \left[ -(k \cdot
  p_1)^3 - 2\, k \cdot p_1\, k \cdot (p_1 - p_2)\, k \cdot p_2 + (k \cdot p_2)^3
  \right. \nonumber \\
&+& \left. \epsilon^2\, \left( \left( (k \cdot p_1)^2 + (k \cdot p_2)^2 \right) +
  6\, k \cdot p_1\, k \cdot p_2 \right) \right] + 8\, m_e^4\, \left[ 3\, \left(
  (k \cdot p_1)^2 + (k \cdot p_2)^2 \right) \right. \nonumber \\
&+& \left. 4\, k \cdot p_1\, k \cdot p_2 \right] + m_\gamma^2\, \left[ -4\, \left(
  (k \cdot p_1)^3 + 6\, \epsilon^4\, k \cdot (p_1 - p_2) - (k \cdot p_2)^3 \right.
  \right. \nonumber \\
&-& \left. 13\, k \cdot p_1\, k \cdot p_2\, k \cdot (p_1 - p_2) + \epsilon^2\,
  \left( -6\, (k \cdot p_1)^2 + 28\, k \cdot p_1\, k \cdot p_2 - 6\, (k \cdot p_2)^2
  \right) \right) \nonumber \\
&+& m_e^2\, \left( 32\, \epsilon^2\, k \cdot (p_1 - p_2) + 12\, \left( (k \cdot
  p_1)^2 + (k \cdot p_2)^2 \right) + 48\, k \cdot p_1\, k \cdot p_2 \right)
  \nonumber \\
&+& \left. 8\, m_e^4\, k \cdot (p_1 - p_2) \right] - 2\, m_\gamma^4\, \left[ 6\,
  \epsilon^4 + 5\, (k \cdot p_1)^2 - 18\, k \cdot p_1\, k \cdot p_2 + 5\, (k \cdot
  p_2)^2 \right. \nonumber \\
&-& \left. 2\, m_e^4 - 8\, \epsilon^2\, \left(2\, k \cdot (p_1 - p_2) + m_e^2
  \right) \right] + 2\, m_\gamma^6\, \left(4\, \epsilon^2 - 3\, k \cdot (p_1 - p_2)
  + m_e^2 \right) \nonumber \\
&-& \left. \left. m_\gamma^8 \right\} \right\}\,\,\, ,\\
\nu_3
&=& 4\, |{\bvec{p}_1}|^2\, |{\bvec{p}_2}|^2\, s_1^2\, s_2^2\, \left\{ \left(\ca^{\!\!\!
  \prime  2} - 2\, \cv^{\!\!\! \prime  2} \right)\, m_e^2\, m_\gamma^2\, \left(2\,
  k \cdot (p_1 - p_2) + m_\gamma^2 \right) \right. \nonumber \\
&+& 12\, \cv^{\!\!\! \prime  2}\, m_e^2\, k \cdot p_1\, k \cdot p_2 - \left(\ca^{\!\!\!
  \prime  2} + \cv^{\!\!\! \prime  2} \right)\, \left\{ 4\, k \cdot p_1\, k \cdot p_2\,
  \left(3\, \epsilon^2 - 2\, k \cdot (p_1 - p_2) \right) \right. \nonumber \\
&-& 2\, m_e^2\, \left( (k \cdot p_1)^2 + (k \cdot p_2)^2 \right) + m_\gamma^2\, \left[
  -6\, \epsilon^2\, k \cdot (p_1 - p_2) + 3\, \left( (k \cdot p_1)^2 + (k \cdot p_2)^2
  \right) \right. \nonumber \\
&-& \left. \left. \left. 14\, k \cdot p_1\, k \cdot p_2 \right] - m_\gamma^4\, \left(
  3\, \epsilon^2 - 4\, k \cdot (p_1 - p_2) \right) + m_\gamma^6 \right\} \right\}\,\,\, ,\\
\nu_4
&=& 4\, \left(\ca^{\!\!\! \prime  2} + \cv^{\!\!\! \prime  2} \right)\, |{\bvec{p}_1}|^3\,
  |{\bvec{p}_2}|^3\, s_1^3\, s_2^3\, \left(2\, k \cdot p_2 - m_\gamma^2 \right)\, \left(
  2\, k \cdot p_1 + m_\gamma^2 \right)
\eea

\section{Electromagnetic excitations in a plasma}
\label{plasmapp}
\setcounter{equation}0

\begin{figure}
\epsfysize=5.0cm
\epsfxsize=10.0cm
\centerline{\epsffile{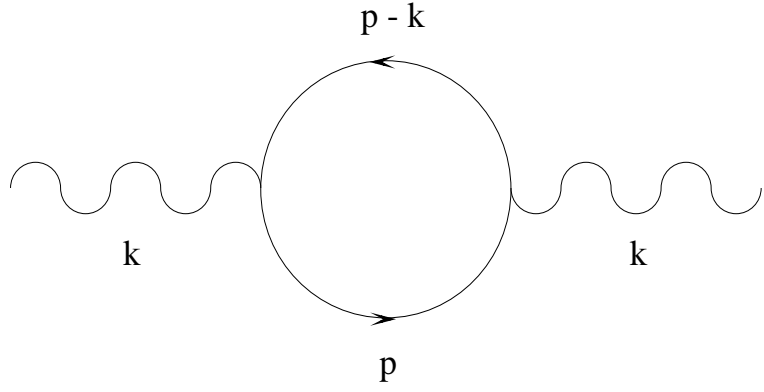}}
\caption{Self-energy diagram for a photon in a plasma.}
\label{pself}
\end{figure}

The equation of motion for the 4-vector potential field $A^\mu$ of a
photon in a plasma can be written as
\be
\left( - k^2 \, g_{\mu \nu} \; + \; \Pi_{\mu \nu} \right) \, A^\nu =
0\,\,\, ,
\label{f19}
\ee
where $\Pi_{\mu \nu}$ is the polarization tensor obtained from the diagram
in Figure \ref{pself} (in the Feynman gauge) \cite{Braaten93},
\cite{Raffelt95}:
\be
\Pi_{\mu\nu} = - 16 \pi\alpha\int\frac{d^3{\bf p}}{(2\pi)^3}\,
\frac{\left[ F_-(E) + F_+(E) \right]}{2 E} \, \frac{p \cdot K\,(K_\mu
p_\nu+K_\nu p_\mu)-K^2 p_\mu p_\nu- (p \cdot K)^2 g_{\mu\nu}} {(p \cdot
K)^2-K^4/4},
\label{Xe}
\ee
with $p=(E,{\bf p})$, $K=(E_\gamma,{\bf k})$ and $E_\gamma =
\sqrt{|{\bvec{k}}|^2+ m_\gamma^2}$, $E= \sqrt{|{\bvec{p}}|^2+ m_e^2}$. It
is useful to express $\Pi_{\mu \nu}$ in terms of form factors by using
Lorentz and gauge invariance, namely $ k^\mu \Pi_{\mu \nu} = k^\nu
\Pi_{\mu \nu} = 0$. The most general form of $\Pi_{\mu \nu}$ (assuming
parity conservation) results to be \cite{Weldon}
\be
\Pi_{\mu \nu} \; = \; \Pi_T \, R_{\mu \nu} \; + \; \Pi_L \, Q_{\mu
\nu}\,\,\, ,
\label{f13}
\ee
where $R_{\mu \nu}$ and $Q_{\mu \nu}$ are given in Eqs.(\ref{rmunu}),
(\ref{qmunu}). The quantities $\Pi_T$ and $\Pi_L$ may be computed from the
relations $\Pi_L = Q^{\mu \nu} \Pi_{\mu \nu}$, $\Pi_T = R^{\mu \nu}
\Pi_{\mu \nu}$ and the results are \footnote{The expressions in
Eqs.(\ref{Pit}) and (\ref{Pil}) are obtained by neglecting the term
$K^4/4$ in the denominator of Eq.(\ref{Xe}). As shown in Ref.
\cite{Braaten93}, this introduces only an error of higher order in
$\alpha$ and, moreover, eliminate the effects of the unphysical process
$\gamma \rightarrow e^+ e^-$, since $\Pi_{T,L}$ remain real-valued at all
temperatures and densities as they should be.}:
\bea
\Pi_T &=& \frac{4 \alpha}{\pi} \int_0^\infty d|{\bvec{p}}| \;
\frac{|{\bvec{p}}|^2}{E} \left( \frac{E_\gamma^2}{|{\bvec{k}}|^2} -
\frac{E_\gamma^2 - |{\bvec{k}}|^2}{|{\bvec{k}}|^2} \frac{E_\gamma}{2 v
|{\bvec{k}}|} \log\left(\frac{E_\gamma + v |{\bvec{k}}|} {E_\gamma - v
|{\bvec{k}}|}\right) \right) \nonumber\\
&\times&\left[ F_-(E) \;+\; F_+(E) \right]\,\,\, , \label{Pit} \\
\Pi_L &=& \frac{4 \alpha}{\pi}~ \frac{E_\gamma^2 - |{\bvec{k}}|^2}
{|{\bvec{k}}|^2} \int_0^\infty d|{\bvec{p}}| \;
\frac{|{\bvec{p}}|^2}{E} \left( \frac{E_\gamma}{v |{\bvec{k}}|} \log
\left(\frac{E_\gamma + v |{\bvec{k}}|} {E_\gamma - v |{\bvec{k}}|}\right)
- 1 - \frac{E_\gamma^2 - |{\bvec{k}}|^2}{E_\gamma^2 - v^2 |{\bvec{k}}|^2}
\right) \nonumber \\
&\times& \left[ F_-(E) \;+\; F_+(E) \right]\,\,\, ,
\label{Pil}
\eea
where $v=|{\bvec{p}}|/E$ is the electron or positron velocity. The
integrals in Eqs.(\ref{Pit}), (\ref{Pil}) over the electron momentum
$|{\bvec{p}}|$ are well approximated \cite{Braaten93,Raffelt95} by the
following expressions that we use in our computations for the neutrino
energy loss rates
\bea
\Pi_T &=& \omega_P^2 \left[ 1+{\frac{1}{2}} G\left(\frac{v_*^2
|{\bvec{k}}|^2}{\omega^2}\right) \right]\,\,\, , \label{Xf1} \\
\Pi_L &=& \omega_P^2 \left[ 1-G\left(\frac{v_*^2
|{\bvec{k}}|^2}{\omega^2}\right) \right] + v_*^2 |{\bvec{k}}|^2
-|{\bvec{k}}|^2\,\,\, ,
\label{Xf2}
\eea
where $v_*\equiv\omega_1/\omega_P$, with the definitions
\bea
\omega_P^2
&\equiv&\frac{4\alpha}{\pi}\int_0^\infty d|{\bvec{p}}|\, \left(
v-{\frac{1}{3}}v^3 \right) \, |{\bvec{p}}| \, \left[ F_-(E) \;+\; F_+(E)
\right]\,\,\, , \\
\omega_1^2 &\equiv&\frac{4\alpha}{\pi}\int_0^\infty d|{\bvec{p}}|\, \left(
{\frac{5}{3}} v^3-v^5 \right) \, |{\bvec{p}}| \,\left[ F_-(E) \;+\; F_+(E)
\right]\,\,\, ,
\label{Xff}
\eea
can be interpreted as a typical velocity of the electrons in the medium.
The function $G$ is defined by
\be
G(x)\equiv\frac{3}{x}\left[1-\frac{2x}{3}-\frac{1-x}{2\sqrt{x}}
\log\left(\frac{1+\sqrt{x}}{1-\sqrt{x}}\right)\right]\,\,\, .
\label{Xfff}
\ee
The dispersion relations $E_{\gamma T,L}(|{\bvec{k}}|)$ for transverse and
longitudinal photon modes are given by the locations of the poles in the
effective photon propagator which, in the Feynman gauge, takes the form
\cite{Weldon}
\be
D_{\mu \nu} = - \ \frac{R_{\mu \nu}}{K^2 - \Pi_T} \, - \, \frac{Q_{\mu
\nu}}{K^2 - \Pi_L}\,\,\, .
\label{propfey}
\ee
The explicit expressions for $E_{\gamma T,L}(|{\bvec{k}}|)$ are then
obtained as the solutions of the implicit equations
\bea
E_{\gamma T,L}^2 \, - \, |{\bvec{k}}|^2 = \Pi_{T,L}(E_{\gamma
T,L},|{\bvec{k}}|)\,\,\, ,
\label{dispph}
\eea
while, near the poles, the scalar parts of the effective propagators are:
\be
\frac{1}{K^2 - \Pi_{T,L}} \; \simeq
\frac{Z_{T,L}}{2 E_{\gamma T,L}} \; \frac{1}{E_\gamma - E_{\gamma T,L}}\,\,\, ,
\ee
with
\be
Z_{T,L} = 1 + \frac{1}{2 E_{\gamma T,L}} \, \left. \frac{\partial
\Pi_{T,L}}{\partial E_\gamma} \right|_{E_\gamma = E_{\gamma T,L}}\,\,\, .
\ee
By inserting Eqs.(\ref{Xf1}), (\ref{Xf2}) into the above equation for
$Z_{T,L}$, we obtain the final expressions for the residue functions used
in the plasmon decay rate \cite{Braaten93}, \cite{Raffelt95}
\bea
Z_{T}&=&\frac{2\, E_{\gamma T}^2\, (E_{\gamma T}^2-v_*^2\,
|{\bvec{k}}|^2)}{E_{\gamma T}^2\, (3\, \omega_P^2 - 2 \, \Pi_T)+(E_{\gamma
T}^2 + |{\bvec{k}}|^2) (E_{\gamma T}^2 - v_*^2\, |{\bvec{k}}|^2)}\,\,\, ,
\label{Xffff1} \\
Z_{L} &=& \frac{2\, E_{\gamma L}^2 (E_{\gamma L}^2-v_*^2\,
|{\bvec{k}}|^2)}{[3\, \omega_P^2 -(E_{\gamma L}^2-v_*^2\,
|{\bvec{k}}|^2)]\, \Pi_L}\,\,\, .
\label{Xffff2}
\eea
From Eq.(\ref{dispph}) we can see that $E_{\gamma T} > |{\bvec{k}}|$ for
all $|{\bvec{k}}|$ while $E_{\gamma L} > |{\bvec{k}}|$ only for
$|{\bvec{k}}| < |{\bvec{k}}|_{\mathrm max}$ with \cite{Braaten93}
\be
|{\bvec{k}}|_{\mathrm max}^2 = \frac{4\, \alpha}{\pi} \int_0^\infty
d|{\bvec{p}}| \; \frac{|{\bvec{p}}|^2}{E} \left( \frac{1}{v}
\log\left(\frac{1 + v}{1 - v}\right) - 1 \right) \left[ F_-(E) + F_+(E)
\right]\,\,\, .
\label{kmax}
\ee
Then longitudinal photon modes can decay into neutrino pairs if their
momentum is lower than the maximum value $|{\bvec{k}}|_{\mathrm max}$.
Note that to the same level of approximation as for Eqs.(\ref{Xf1}),
(\ref{Xf2}) we have \cite{Braaten93}
\be
|{\bvec{k}}|_{\mathrm max} =
\omega_P\,\left[\frac{3}{v_*^2}\left(\frac{1}{2v_*}
\log\left(\frac{1+v_*}{1-v_*}\right)-1\right)\right]^{1/2}
\label{Xg}\,\,\, .
\ee

\end{document}